\documentclass{emulateapj}

\newcommand{\citepeg}[1]{\citep[{e.g.,}][]{#1}}
\def\Swift{{\textit{Swift}}\,}

\shorttitle{The Afterglow of GRB 130427A}
\shortauthors{Perley et al.}

\begin{document}

\title{The Afterglow of GRB 130427A from $1$ to $10^{16}$ GHz}

\def\cit{1}
\def\hubble{2}
\def\gsfc{3}
\def\jssi{4}
\def\berkeley{5}
\def\gwu{6}
\def\leicester{7}
\def\warwick{8}
\def\mpe{9}
\def\tautenburg{10}
\def\adiyaman{11}
\def\yunnan{12}
\def\klsecb{13}
\def\iram{14}
\def\iaa{15}
\def\uadi{16}
\def\nrao{17}
\def\stsci{18}
\def\sabanco{19}
\def\istanbul{20}
\def\riken{21}
\def\sri{22}
\def\ralcao{23}

\def\mail{*}

\author{D.~A.~Perley\altaffilmark{\cit,\hubble,\mail}, 
        S.~B.~Cenko\altaffilmark{\gsfc,\jssi,\berkeley},
        A.~Corsi\altaffilmark{\gwu},
        N.~R.~Tanvir\altaffilmark{\leicester},
        A.~J.~Levan\altaffilmark{\warwick},
        D.~A.~Kann\altaffilmark{\mpe,\tautenburg},
        E.~Sonbas\altaffilmark{\adiyaman},
        K.~Wiersema\altaffilmark{\leicester},
        W.~Zheng\altaffilmark{\berkeley}, 
        X.-H.~Zhao\altaffilmark{\yunnan,\klsecb},
        J.-M.~Bai\altaffilmark{\yunnan,\klsecb},
        M.~Bremer\altaffilmark{\iram},
        A.~J.~Castro-Tirado\altaffilmark{\iaa,\uadi},
        L.~Chang\altaffilmark{\yunnan,\klsecb}, 
        K.~I.~Clubb\altaffilmark{\berkeley},
        D.~Frail\altaffilmark{\nrao},
        A.~Fruchter\altaffilmark{\stsci},
        E.~G\"o\u{g}\"u\c{s}\altaffilmark{\sabanco},
        J.~Greiner\altaffilmark{\mpe},
        T.~G\"uver\altaffilmark{\istanbul},
        A.~Horesh\altaffilmark{\cit},
        A.~V.~Filippenko\altaffilmark{\berkeley},
        S.~Klose\altaffilmark{\tautenburg},
        J.~Mao\altaffilmark{\yunnan,\riken},
        A.~N.~Morgan\altaffilmark{\berkeley},
        A.~S.~Pozanenko\altaffilmark{\sri},
        S.~Schmidl\altaffilmark{\tautenburg},
        B.~Stecklum\altaffilmark{\tautenburg},
        M.~Tanga\altaffilmark{\mpe},
        A.~A.~Volnova\altaffilmark{\sri},
        A.~E.~Volvach\altaffilmark{\ralcao},
        J.-G.~Wang\altaffilmark{\yunnan,\klsecb}, 
        J.-M.~Winters\altaffilmark{\iram}, and
        Y.-X.~Xin\altaffilmark{\yunnan,\klsecb}
}

\altaffiltext{\cit}{Department of Astronomy, California Institute of Technology,
MC 249-17,
1200 East California Blvd,
Pasadena CA 91125, USA}
\altaffiltext{\hubble}{Hubble Fellow}
\altaffiltext{\gsfc}{Astrophysics Science Division, NASA Goddard Space Flight Center, Mail Code 661, Greenbelt, MD 20771, USA}
\altaffiltext{\jssi}{Joint Space Science Institute, University of Maryland, College Park, Maryland 20742, USA}
\altaffiltext{\berkeley}{Department of Astronomy, University of California, Berkeley, CA 94720-3411, USA}
\altaffiltext{\gwu}{Physics Department, George Washington University, 725 21st St, NW Washington, DC 20052, USA}
\altaffiltext{\leicester}{Department of Physics and Astronomy, University of Leicester, Leicester LE1 7RH, UK}
\altaffiltext{\warwick}{Department of Physics, University of Warwick, Coventry CV4 7AL, UK}
\altaffiltext{\mpe}{Max-Planck-Institut f\"ur extraterrestrische Physik, Giessenbachstra\ss e, 85748 Garching, Germany}
\altaffiltext{\tautenburg}{Th\"uringer Landessternwarte Tautenburg, Sternwarte 5, 07778 Tautenburg, Germany}
\altaffiltext{\adiyaman}{Department of Physics, University of Adiyaman, 02040 Adiyaman, Turkey}
\altaffiltext{\yunnan}{Yunnan Observatories, Chinese Academy of Sciences, P.O. Box 110, 650011 Kunming, China}
\altaffiltext{\klsecb}{Key Laboratory for the Structure and Evolution of Celestial Bodies, Chinese Academy of Sciences, P.O. Box 110, 650011 Kunming, China}
\altaffiltext{\iram}{Institute de Radioastronomie Millim\`etrique (IRAM), 300 rue de la Piscine, 38406 Saint Martin d\' \rm H\`eres, France.}
\altaffiltext{\iaa}{Instituto de Astrof\'{\i}sica de Andaluc\'{\i}a (IAA-CSIC), Glorieta de la Astronom\'{\i}a s/n, E-18008 Granada, Spain}
\altaffiltext{\uadi}{Unidad Asociada Departamento de Ingenier\'{\i}a de Sistemas y Autom\'atica, E.T.S. de Ingenieros Industriales, Universidad de M\'alaga, Spain.}
\altaffiltext{\nrao}{National Radio Astronomy Observatory, P.O. Box O, Socorro, NM 87801, USA}
\altaffiltext{\stsci}{Space Telescope Science Institute, Baltimore, MD 21218, USA}
\altaffiltext{\sabanco}{Sabanc\i~University, Orhanl\i - Tuzla, \.Istanbul 34956 Turkey}
\altaffiltext{\istanbul}{Istanbul University Science Faculty, Department of Astronomy and Space Sciences, 34119, University-Istanbul, Turkey}
\altaffiltext{\riken}{Astrophysical Big Bang Laboratory, RIKEN, Saitama 351-0198, Japan}
\altaffiltext{\sri}{Space Research Institute, 117997, 84/32 Profsoyuznaya, Moscow, Russia}
\altaffiltext{\ralcao}{Radio Astronomy Laboratory of the Crimean Astrophysical Observatory, Katsiveli, Yalta 98688, Ukraine}

\altaffiltext{\mail}{e-mail: dperley@astro.caltech.edu }

\slugcomment{Submitted to ApJ 2013-07-17, accepted 2013-11-18}

\begin{abstract}
We present multiwavelength observations of the afterglow of GRB\,130427A, the brightest (in total fluence) gamma-ray burst of the past 29 years.  Optical spectroscopy from Gemini-North reveals the redshift of the GRB to be $z = 0.340$, indicating that its unprecedented brightness is primarily the result of its relatively close proximity to Earth; the intrinsic luminosities of both the GRB and its afterglow are not extreme in comparison to other bright GRBs.   We present a large suite of multiwavelength observations spanning from 300~s to 130~d after the burst and demonstrate that the afterglow shows relatively simple, smooth evolution at all frequencies, with no significant late-time flaring or rebrightening activity.  The entire dataset from 1~GHz to 10~GeV can be modeled as synchrotron emission from a combination of reverse and forward shocks in good agreement with the standard afterglow model, providing strong support to the applicability of the underlying theory and clarifying the nature of the GeV emission observed to last for minutes to hours following other very bright GRBs.  A tenuous, wind-stratified circumburst density profile is required by the observations, suggesting a massive-star progenitor with a low mass-loss rate, perhaps due to low metallicity.  GRBs similar in nature to GRB\,130427A, inhabiting low-density media and exhibiting strong reverse shocks, are probably not uncommon but may have been difficult to recognize in the past owing to their relatively faint late-time radio emission; more such events should be found in abundance by the new generation of sensitive radio and millimeter instruments.
\end{abstract}

\section{Introduction}
\label{sec:intro}

The majority of long-duration gamma-ray bursts (GRBs) detected by orbiting satellites are extremely energetic events originating from the distant universe:  the mean redshift among \Swift GRBs is $z \approx 2.0$, and $\sim 80$\% of \Swift events originate from $z>1$ \citepeg{Jakobsson+2006,Jakobsson+2012,Fynbo+2009}.  While this makes GRBs excellent potential probes of early phases of cosmic history, it also implies that nearby analogs of these high-redshift events must be relatively rare: the ratio of observable comoving volume within the range $0 < z < 0.4$ compared to $1 < z < 3$, for example, is approximately a factor\footnote{Here and elsewhere we assume a standard $\Lambda$CDM cosmological model with $\Omega_\Lambda=0.7$, $\Omega_m = 0.3$, $h = 0.7$.} of 60.  Because GRBs are associated with star formation, the sharp decline in the cosmic star-formation rate since $z \approx 1$ (by a factor of 5--10; e.g., \citealt{Madau+1998}) further serves to reduce the relative fraction of GRBs observed from the nearby universe.  The probable sensitivity of the GRB rate to metallicity (e.g., \citealt{LeFloch+2006,Modjaz+2008,Kistler+2008,Butler+2010,Levesque+2010g,Graham+2013,Robertson+2012}; cf. \citealt{Savaglio+2009,Mannucci+2011,Elliott+2012}) also decreases the local rate.

Indeed, a simple scaling of the observed $z \approx 1$--2 GRB rates would naively suggest that ``nearby'' events (those at $z<0.4$) should be extraordinarily uncommon, perhaps one per decade within {\it Swift}'s field of view.  Fortunately, however, GRBs span a wide range of luminosities \citep{Butler+2010,Cao+2011} and a population of less luminous but intrinsically more common events that cannot be detected at higher redshifts becomes visible in the nearby universe \citep{Cobb+2006,Guetta+2007,Soderberg+2006}, raising the observed rate of nearby GRBs to a more respectable (but still relatively low) $\sim1$~yr$^{-1}$ during the \Swift era.  The existence of this population has been critical in tying GRBs conclusively to massive stellar death, since at $z<0.4$ optical observations are capable of unambiguously recognizing an accompanying supernova (SN) signature and classifying it spectroscopically (see \citealt{Woosley+2006} for a review), whereas at high redshifts this task is very challenging or impossible.

\begin{figure*}
\centerline{
\includegraphics[scale=0.75,angle=0]{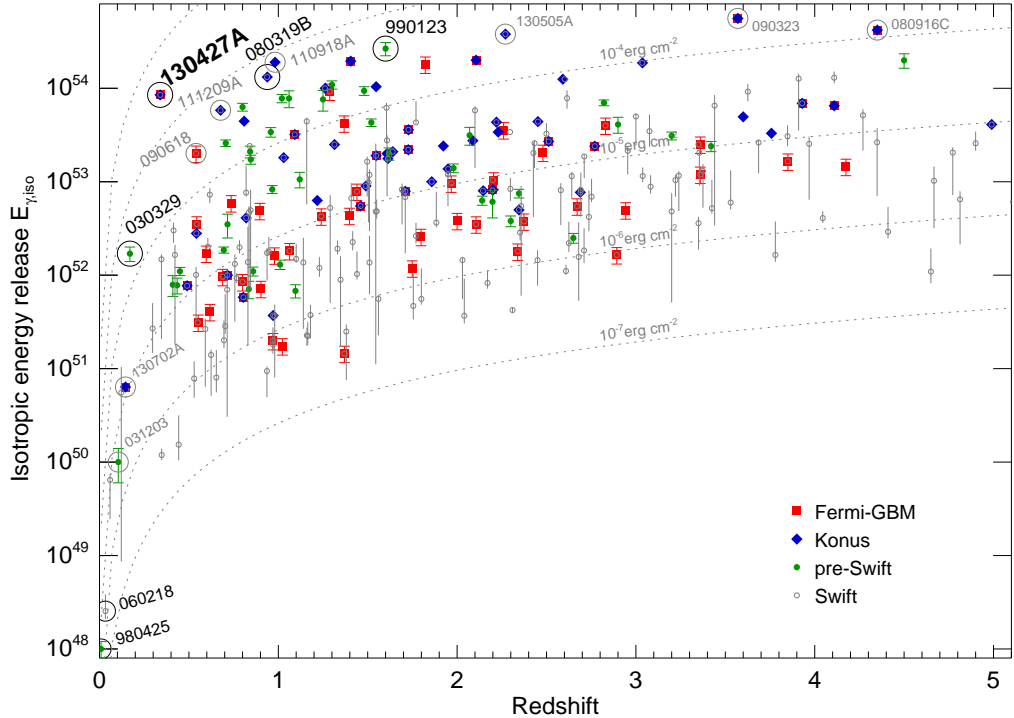}}
\caption{Total (bolometric) isotropic-equivalent gamma-ray energy release of pre-\Swift, \Swift, and \emph{Fermi}-GBM long-duration GRBs versus redshift.  ``High-luminosity'' ($E_{\gamma,{\rm iso}} \gtrsim 10^{52}$ erg) GRBs dominate the observed population and represent the only type of GRB visible at $z>2$.  However, such events have an extremely low intrinsic rate and are rarely observed in the nearby universe due to simple volumetric considerations. Studies of low-redshift GRBs have instead been forced to target more intrinsically common populations of low-luminosity GRBs that may not serve as good analogs of the energetic, high-$z$ population.  GRB\,130427A, the subject of this paper, is the closest example by far of a highly luminous GRB.  Dotted curves are lines of constant fluence. The bottom-right portion of the diagram is empty owing to the undetectability of low-luminosity bursts beyond very low redshifts.
($E_{\gamma,{\rm iso}}$ values are taken from \citealt{Amati+2006}, \citealt{Goldstein+2012},\citealt{Paciesas+2012}, and \citealt{Butler+2007b}, and from Konus GCN Circulars: \citealt{GCN4030,GCN4150,GCN4238,GCN4599,GCN4989,GCN5460,GCN5722,GCN5837,GCN5984,GCN6049,
GCN6960,GCN7482,GCN7487,GCN7589,GCN7812,GCN7854,GCN7862,GCN7995,GCN8548,
GCN8721,GCN8776,GCN9030,GCN9959,GCN10045,GCN10057,
GCN10083,GCN10209,GCN10590,GCN10882,GCN11127,GCN11251,GCN11723,GCN12008,GCN12166,
GCN12362,GCN12433,GCN12663,GCN12790,GCN12872,GCN13446,GCN13736,
GCN14010,GCN14368,GCN14417,GCN14575,GCN14808,GCN9196,GCN9821,GCN14702,GCN11697,GCN9422,GCN12276}.) Some GRBs of particular interest (very luminous and nearby events) are circled and labeled.}
\label{fig:eiso}
\end{figure*}

However, the differences in intrinsic rate and luminosity between these ``nearby'' events and the high-redshift population are quite large. 
The typical GRB selected by \Swift or other major satellites has an isotropic-equivalent energy scale of $E_{\gamma,{\rm iso}} \approx 10^{52}$--$10^{53}$ erg, which is about the energy scale necessary for detection at $z \approx 1$ (Fig. \ref{fig:eiso}).
In contrast, the two nearest \Swift GRBs (060218 at $z=0.033$ and 100316D at $z=0.059$) and the two nearest pre-\Swift GRBs (980425 at $z=0.0085$ and 031203 at $z=0.105$) produced only $E_{\gamma,{\rm iso}} \approx 10^{48}$--$10^{49}$ erg, a difference of four orders of magnitude.   These nearby events couple very little or no energy to the highly relativistic emission normally responsible for producing a GRB \citep{Kaneko+2007}, show no evidence for collimation \citep{Kulkarni+1998,Soderberg+2004,Soderberg+2006}, and early X-ray/UV/optical observations reveal an expanding thermal component instead of a classical optical/X-ray afterglow \citep{Campana+2006,Starling+2011}\footnote{This thermal component may exist in ``standard'', high-luminosity GRBs as well, but is subdominant relative to the afterglow \citep{Sparre+2012}.}.   This is probably because the fastest ejecta in these events do not contain sufficient energy to produce a bright relativistic shock wave in their surrounding media (as universally seen in high-luminosity GRBs), so at most times and frequencies the shock is dominated by other emission processes such as shock breakout and the SN itself, precluding the use of these events for studies of afterglow emission.  

Until now, the best nearby analog of a traditional high-luminosity GRB has been GRB 030329 at $z=0.169$.   With $L_{\rm iso} \approx 10^{51}$ erg s$^{-1}$, it would likely have been detected (by \emph{Swift}) as far away as $z \approx 2$; this event had an extremely bright and well-studied optical/millimeter/X-ray afterglow as well as a spectroscopically confirmed SN that emerged after a few days \citep{Price+2003,Tiengo+2003,Greiner+2003,Sheth+2003,Hjorth+2003,Stanek+2003}.   However, until 2013 this event has remained singular:  no other comparably luminous GRB has been found at $z<0.4$.  Even GRB 030329 is at the low end of the overall GRB population in terms of its gamma-ray energetics, and was peculiar in many ways: in particular, the optical light curve showed continued variability and rebrightenings as late as $\sim8$~days post-trigger \citep{Uemura+2003,Lipkin+2004}, and its bright and long-lived radio afterglow seemed to require a second, wide jet unassociated with any gamma-ray emission \citep{Berger+2003}.  

In this paper, we present observations of \textit{Swift}/\textit{Fermi} GRB\,130427A, the closest high-luminosity ($E_{\gamma,{\rm iso}} > 10^{51}$ erg) GRB since GRB 030329.  With $z=0.34$ and $E_{\gamma,{\rm iso}} \approx 8 \times 10^{53}$ erg, its combination of proximity and luminosity are unprecedented in the history of the field, producing the highest $\gamma$-ray and X-ray fluence of any GRB or afterglow observed during the past 29 years.  Furthermore, this burst occurred under a series of favorable circumstances for observations.  The GRB position was located within the field of view of the {\it Fermi} Large Area Telescope (LAT; \citealt{Atwood+2009}), providing coverage of the GeV-photon component at early times and continuing for many hours after the GRB.  It also occurred over the western hemisphere near local midnight during a period of good weather in the western United States, enabling a number of telescopes to observe the GRB at optical wavelengths within minutes, or in one case starting before the burst began \citep{GCN14476}.

For all these reasons, GRB\,130427A is a keystone event that is likely to represent a gold standard for comparisons with other GRB afterglows for decades.  In this paper, we present a large suite of multiwavelength observations of the afterglow of GRB\,130427A stretching from the radio band to X-rays and from three minutes to four months after the burst. Our acquisition and reduction of the observational data is presented in \S \ref{sec:obs}.  Examination of the key features of the observations as a function of wavelength, including detailed comparisons to samples of past GRBs, is presented in \S \ref{sec:behavior}.  Having identified the key observational features, in \S \ref{sec:modeling} we then attempt to explain the data using a standard reverse+forward shock synchrotron model.  We find that this model provides an excellent description of the entire dataset from 400~s to 130~days and at frequencies ranging from the low-frequency radio to the high-energy gamma-rays, providing support for the standard afterglow model and explaining the origin of the long-lived LAT emission seen in this and previous GRBs as a simple extension of the forward shock.  We summarize our conclusions in \S \ref{sec:conclusions} and examine the implications of our results for modeling of more complex GRBs and for the GRB progenitor.

\section{Observations}
\label{sec:obs}

\subsection{\Swift BAT and XRT}
\label{sec:batxrt}

GRB\,130427A triggered the Burst Alert Telescope (BAT; \citealt{Barthelmy+2005}) on board the \Swift satellite \citep{Gehrels+2004} on 2013 April 27 at 07:47:57 (UT dates are used throughout this paper).  This trigger time actually corresponds to a point near the end of burst activity, as the BAT was in the middle of a preplanned slew when burst emission began and could not trigger until the slew was complete.  Consequently, the BAT trigger time does not provide a useful reference time for the burst; we instead employ the {\it Fermi} Gamma-Ray Burst Monitor (GBM; \citealt{Meegan+2009}) trigger time of 07:47:06.42 \citep{GCN14473} as $t_0$ in all of our subsequent analyses, an adjustment of 50.58~s for times referenced to the BAT trigger.

Following the end of its preplanned slew and trigger, \Swift slewed immediately to the BAT location and began observations with the X-Ray Telescope (XRT; \citealt{Burrows+2005}) beginning at 07:50:17.7 ($t-t_0 = 190.8$~s).  Observations continued until $t-t_0 = 1984$ s, at which point \Swift slewed to another location owing to orbital visibility constraints.  After a gap of about 20~ksec (0.23~d), \Swift returned to the source for further observations; regular additional observing epochs continued as long as the position remained visible to \emph{Swift}.

We downloaded the \Swift BAT and XRT light curves and spectral analysis from the \emph{Swift} Burst Analyser \citep{Evans+2010}\footnote{http://www.swift.ac.uk/burst\_analyser/} and \Swift XRT repository \citep{Evans+2007,Evans+2009}.  Specifically, we obtained the 15--50~keV BAT flux light curve (in 64~ms, 1~s, 10~s, and signal-to-noise ratio [S/N] = 7 binning modes) and the 0.3--10~keV XRT flux light curve; each light curve was appropriately rebinned by our own scripts depending on the application.  Where necessary, these bandpass-integrated fluxes were converted to flux-density values in Jy using a smoothed value of the photon index $\Gamma$ for each bin and a correction factor of 1.16 for X-ray absorption (taken from the ratio of absorbed to unabsorbed fluxes on the XRT spectral analysis page for this event\footnote{http://www.swift.ac.uk/xrt\_spectra/00554620/}).

\subsection{UVOT}
\label{sec:uvot}

\begin{figure*}
\centerline{
\includegraphics[scale=0.35,angle=0]{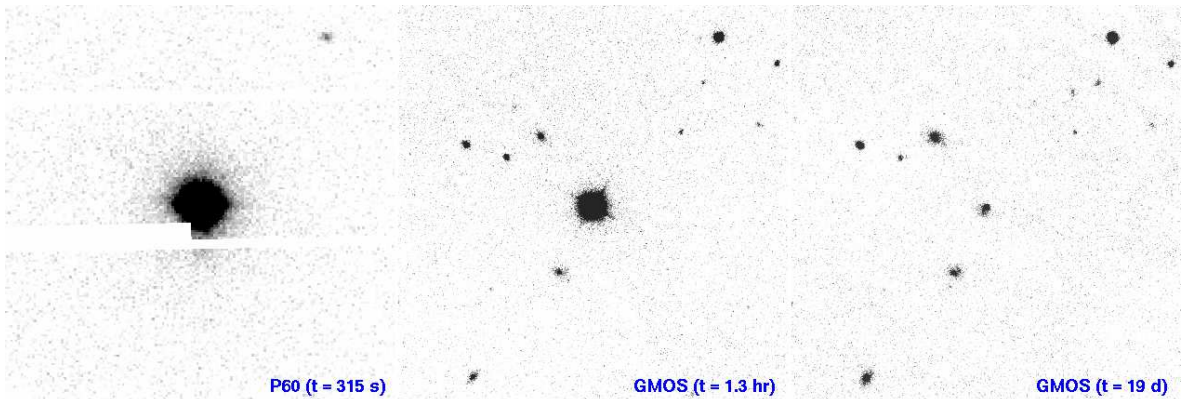}}  
\caption{Afterglow discovery image from the P60 (left panel) taken starting at 315~s post-trigger, Gemini-North acquisition image taken at $t=1.3$ hours (middle panel) at the time of the first epoch of spectroscopy identifying the redshift to be 0.340, and an additional Gemini-North image taken at 19~d (right panel), after the afterglow had faded below $r=22$ and light was dominated by the supernova and extended host galaxy.   The white region on the left panel is a region of of bad columns on the P60 CCD.  North is up and east is to the left; the field is approximately 1\arcmin\ in diameter.}
\label{fig:imfig}
\end{figure*}

Ultraviolet-Optical Telescope (UVOT) data were also acquired by \Swift in parallel with XRT follow-up observations beginning at $t-t_0 = 197$ s.  The GRB field is at high Galactic latitude ($b = +72^{\circ}$) and there are relatively few bright stars nearby; consequently, the spacecraft experienced difficulties guiding and most of the initial exposures are trailed, though this difficulty was corrected in subsequent epochs.

We reduced the UVOT data using standard procedures within the HEASoft\footnote{http://heasarc.nasa.gov/lheasoft} environment (e.g., \citealt{Brown+2009}).  Flux from the transient was extracted from a 3\arcsec\ radius aperture for images with good star lock (a much larger aperture was used on trailed exposures to include all of the trailed flux), with a correction applied to put the photometry on the standard UVOT system \citep{Poole+2008}.  For observations after $t=8$ days the object is not detected in individual epochs, so we stacked observations in three blocks spanning the time periods $t=9$--15 d, 16--30 d, and 30--60 d.  The resulting measurements are listed in Table~\ref{tab:optphot}.

\subsection{Palomar 60 inch Telescope}
\label{sec:p60}

The Palomar 60 inch telescope (P60; \citealt{Cenko+2006}) responded automatically to the BAT trigger and began imaging the field starting at 07:52:21.7 (Fig. \ref{fig:imfig}), detecting a bright source at the location reported by \cite{GCN14451}.  This initial set of observations consisted of a repeating cycle of 60~s exposures in the $r$, $i$, and $z$ filters.  P60 temporarily slewed away from the target after completing this sequence $\sim 90$~min later, but was manually instructed to return to the field and continue observations, this time in a repeating cycle of 60 s exposures in the $g$, $r$, and $i$ filters.  Observations continued until a telescope limit was reached at airmass 4.2.  P60 was not available for observations the following night, but the GRB was monitored the night after that (and for a majority of the next several nights) in the $g$, $r$, $i$, and $z$ filters for most of the window in which it was observable, switching to 120~s and then 180~s exposures.  As the source faded and the Moon brightened, the $z$ and $g$ exposures were dropped in favor of $r$ and $i$; observations continued nightly (except during periods of bad weather) until May 31, after which a less regular cadence was used.

Basic reductions (bias subtraction, flat-fielding, and astrometry) are provided in real time by an automated IRAF\footnote{IRAF is distributed by the National Optical Astronomy Observatory, which is operated by the Association of Universities for Research in Astronomy (AURA) under cooperative agreement with the National Science Foundation (NSF).} pipeline.  During periods when the Moon was up significant scattered moonlight was visible in the frames; we subtract it using a scaled model of the moonlight pattern, which is approximately constant with phase.  In addition, $z$-band frames are significantly fringed and $i$-band frames very marginally fringed; we create a fringe frame from observations taken during dark time and subtract it to remove this pattern.  Observations are then stacked using SWarp\footnote{http://www.astromatic.net/software/swarp} to improve the S/N in observations taken after the first night.  

For observations taken on the night of the burst, the host-galaxy flux is insignificant ($<0.5$\% of the total flux) and we performed point-spread function (PSF) matched photometry of the afterglow directly on individual P60 frames using the \texttt{DAOPHOT} package.  The afterglow mildly saturated the first few frames; we include these data by fitting only to the nonsaturated pixels.  Pixels affected by bad columns (e.g., Fig.\ \ref{fig:imfig}) were also excluded.  The resulting photometry was calibrated with respect to nearby point sources from Data Release 9 (DR9; \citealt{DR9}) of the Sloan Digital Sky Survey (SDSS) and is in the AB system \citep{Oke1974}.  At later times ($t \gtrsim 1$~d), the contribution from the underlying host-galaxy light at the afterglow location is no longer negligible.  To directly measure the afterglow brightness, we subtracted reference SDSS frames from our P60 imaging using the publicly available High Order Transform of PSF ANd Template Subtraction (HOTPANTS\footnote{See http://www.astro.washington.edu/users/becker/hotpants.html}).  The brightness of the transient was measured in the resulting subtracted frames again using \texttt{DAOPHOT}.  (We note that this procedure differs from the technique of simply subtracting the galaxy in flux space employed for all other instruments; see \S \ref{sec:host}.)  Results are reported in Table \ref{tab:optphot}; for consistency with the magnitudes from other instruments reported in this table (which are not host-subtracted), our estimate of the total host flux (\S \ref{sec:host}) is re-added to the magnitude column.

\subsection{Gemini-North}
\label{sec:gemini}

Shortly following the discovery of GRB\,130427A and its afterglow, we initiated observations with the Gemini North telescope on Mauna Kea.  A spectroscopic sequence of four 400~s exposures began at 09:19 (91 min after the GRB), using the B600 grating and covering a wavelength range of 3824--6707~\AA.  A second sequence of three 300~s exposures was obtained starting at 10:44 (177~min after the GRB), again utilizing the B600 grating but set to a bluer central wavelength for a coverage of 3080--5955~\AA\ (but the sensitivity is poor shortward of $\sim 3300$~\AA).

\begin{figure}
\centerline{
\includegraphics[scale=0.52,angle=0]{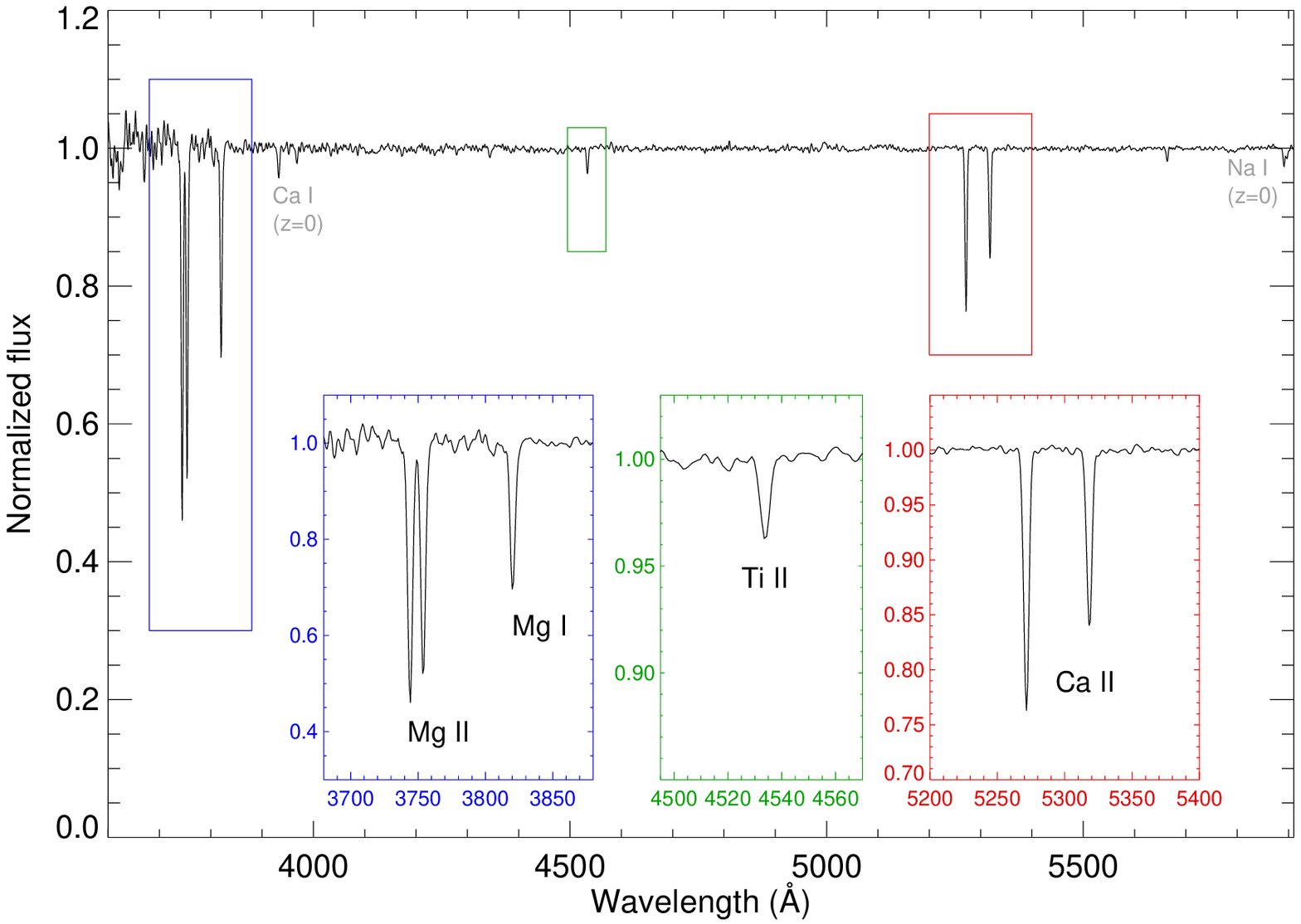}} 
\caption{Early-time Gemini-North spectroscopy of the afterglow of GRB\,130427A.  The spectrum (normalized here to 1.0 based on a polynomial fit to the continuum; photometric observations of the afterglow during this period show that the true continuum shape is a simple power law of $F_\nu \propto \nu^{-0.4}$ after extinction correction) is almost entirely featureless with the exception of a small number of metal lines at a common redshift of 0.340, which we associate with the GRB and its host galaxy.}
\label{fig:spectrum}
\end{figure}

The data were reduced and combined in IRAF via the usual GMOS pipeline. No standard star was observed on the same night, so we plot in Figure~\ref{fig:spectrum} the normalized spectrum. We clearly identify lines of Ca~II, Mg~II, and Mg~I, providing the first measurement of the redshift of GRB~130427A of $z=0.340$ \citep{GCN14455}. Despite the exceptionally high S/N (the afterglow was at $R = 14.9$--15.3 mag during the observations), the spectrum is essentially featureless outside these transitions, with the exception of a weak detection of Ti~II $\lambda$3384, Galactic ($z=0$) Ca~II and Na~I, and a possible detection of Mn~II in the ultraviolet (UV) at low S/N.  This is not unexpected; at $z=0.340$ the strongest metal lines remain in the UV and lie blueward of our spectral range.   The overall spectrum is broadly comparable to the composite long-duration GRB spectrum of \cite{Christensen+2011} at these wavelengths.

The spectrum also shows some weak Galactic features.  The S/N and resolution of the GMOS spectra are sufficient to fit the two components of the Galactic Na\,{\sc I} D doublet absorption feature (the D$_1$ and D$_2$ components). We fit Gaussian 
functions to the two lines in the first GMOS spectrum, which has the highest continuum S/N
in the region of interest. We find a summed equivalent width of the two components of 
EW(D$_1$ + D$_2$) =  $0.193 \pm 0.017$ \AA. Using the empirical relation between sodium absorption
strength and dust extinction as derived by \citealt{Poznanski+2012} (their Equation 9), this implies
$E_{B-V} = 0.024$ mag. This is in agreement with the value as derived from the \cite{SFD} maps, but we note that at
the low resolution of GMOS there may be a component of atmospheric sodium contaminating the 
measurement.

Follow-up observations of the associated SN were conducted on 2013 May 16 and 17.  While the primary motivation of both epochs was for spectroscopy of the SN (which we do not discuss here; a detailed multi-epoch study of the SN spectroscopic properties will be left for future work), both observations were preceded by a short imaging acquisition sequence.  Specifically, on the first night a four-filter sequence of $griz$ was employed, but the second-night acquisition was performed with only the $r$-band filter.  We reduce these imaging observations using the Gemini reduction tools in IRAF, and measure the magnitude of the host+afterglow using a 3.0$\arcsec$ radius error circle.  

\subsection{UKIRT}
\label{sec:ukirt}

We observed the field of GRB\,130427A with the United Kingdom Infrared Telescope (UKIRT) using the Wide Field Camera (WFCAM) in the $JHK$ filters at a range of epochs from a few hours to a few weeks post-burst.  The images were pipeline processed by the Cambridge Astronomical Survey Unit (CASU).  Aperture photometry was performed using the {\sc Gaia} software, and calibrated relative to seven 2MASS stars in the field (each having typical quoted errors of 0.02--0.03 mag, leading to a zero-point statistical uncertainty of about 0.01 mag).   We extracted the photometry with both 1.5\arcsec\ and 3.0\arcsec\ radii apertures and use a constant flux offset (calculated from the median flux difference between the two radii for each filter) to correct the 1.5\arcsec\ aperture photometry for the contribution of the outer parts of the host galaxy.  The final magnitudes (corresponding to an effective 3.0\arcsec\ radius aperture and including all of the host flux) are reported in Table \ref{tab:optphot}.

\subsection{GMG}

Imaging observations were carried out each night between 2013 April 27 and April 30 with the Yunnan Faint Object Spectrograph and Camera (YFOSC) instrument of the Lijiang 2.4~m Gao-Mei-Gu (GMG) telescope in Yunnan Observatories, China.  Data were taken with the $Briz$ filters\footnote{In a preliminary report to the GCN Circulars \citep{GCN14466} the $B$-band observation was incorrectly reported as $g$ band.} on the first night following the GRB and with the $r$ filter exclusively during the three subsequent nights. Bias-correction and flat-fielding were performed using standard IRAF routines and the magnitude of the transient was measured with respect to SDSS comparison stars using a 3$\arcsec$ radius aperture.   The S/N of each individual image is high ($\sim50$), and the uncertainty of most observations is dominated by the calibration to the standard stars.

\subsection{T100}
\label{sec:t100}

Images of GRB\,130427A were also acquired with the 1.0 m Telescope (T100) at TUBITAK National Observatory (TUG) in Turkey. We obtained six sets of $R$-band observations on April 27.75, 28.73, 29.77, and 30.73, and on May 1.80 and 2.79. Initial reduction (bias subtraction and flat-fielding) was performed with the European Southern Observatory Munich Image Data Analysis System (ESO-MIDAS) software environment (version 12FEBpl1.3) and IDL codes.  Significant variations are observed in sky brightness across the chip even after flat-fielding using standard reference flats, so we further flat-fielded the observations by median-combining a series of exposures from the first night of observations and dividing all images by the resulting super-sky flat; observations were stacked as appropriate to improve the S/N per image.  Aperture photometry on all images was performed with a custom IDL routine using SDSS field stars as calibrators and an aperture radius of 3\arcsec.

\subsection{GROND}

Imaging was acquired with the simultaneous seven-color Gamma-
Ray burst Optical and Near-infrared Detector (GROND; \citealt{Greiner+2008} )
mounted on the 2.2~m MPG/ESO telescope at 
La Silla Observatory, Chile. We observed the field on 2013 April 28, 
April 29, May 11, May 13, and May 19, the long hiatus being caused by 
detector downtime. Owing to the northerly declination of the 
source, all observations were obtained at moderate to high airmass 
and bad seeing conditions. Data reduction was performed via a custom 
pipeline calling upon IRAF routines, comparable to the methods 
described in more detail by \cite{Kruehler+2009} and 
\cite{Yoldas+2008}. During the first night, the afterglow is still 
very bright and easily detected in single exposures, but due to the 
sparse field and bad observing conditions, data had to be stacked to 
allow astrometric and photometric analysis. Calibration was performed against 
SDSS (optical) and 2MASS near-infrared (NIR) field stars for the late (SN) 
epochs, and the resulting catalogs for these images were used as input 
catalogs to astrometrize the early images. Photometry was then 
performed versus catalogs of reference stars derived from higher-quality 
P60 ($griz$) and UKIRT ($JHK$) images. Early measurements were obtained 
using PSF photometry in the optical and seeing-matched aperture 
photometry in the NIR, whereas SN-epoch measurements were obtained using a 
fixed 3\arcsec\ radius aperture in all bands to encompass the entire host 
galaxy.

\subsection{Lick}
\label{sec:lick}

GRB\,130427A was also observed with both the 0.76~m Katzman Automatic Imaging Telescope \citep[KAIT;][]{Filippenko+2001,Li+2003}, the 1\,m Nickel Telescope, and the 3\,m Shane telescope at Lick Observatory. KAIT and Nickel observations  were performed in the $BVRI$ filters from one to several days after the burst when the afterglow was still bright,  while late-time Shane data were in $BVR$ using the Kast spectrograph \citep{Miller+1993} in imaging mode. We performed photometry on the images with a 2$\farcs$5 radius aperture and calibrated to nearby SDSS stars, transformed to $BVRI$ using the equations of Lupton\footnote{http://www.sdss.org/dr7/algorithms/sdssUBVRITransform.html}.  Both KAIT and Kast have relatively small fields of view, and only a small number of comparison stars (2--3 for KAIT and 1 for Kast) were available for calibration, possibly producing a small amount of additional systematic uncertainty in the photometry.

\subsection{Palomar 200 inch Hale Telescope}
\label{sec:p200}

We observed the location of GRB\,130427A with the Large-Format Camera (LFC) on the Palomar 200 inch telescope on 2013 May 5.   We acquired five dithered exposures in the $g$ band and four in $r$. The data were reduced with a Python code written by B. Sesar.  Photometry was performed on each image individually in IDL using a 3$\arcsec$ radius aperture, and individual exposures were averaged together in flux space for each filter.

\subsection{Tautenburg}
\label{sec:tautenburg}

We obtained two epochs of observations on May 5 and May 12 using the  1.34\,m Schmidt telescope of the Th\"uringer Landessternwarte Tautenburg, Germany. Images were acquired using the 4k camera in the first and the 2k camera in the second epoch. Photometry was calibrated against SDSS field stars transformed to $R_C$ magnitudes using the transformations of Lupton. Magnitudes were derived using a fixed 3\arcsec\ radius aperture to encompass the entire host galaxy. 

\subsection{Keck}
\label{sec:keck}

Images of GRB\,130427A were taken using the Low Resolution Imaging Spectrometer \citep[LRIS;][]{Oke1995} on 2013 May 9 in the $g$ and $V$ filters
and on 2013 May 10 in the $g$ and $R$ filters.  These data were reduced via standard techniques using a custom IDL pipeline.  Further imaging was conducted with the Deep Imaging Multi-Object Spectrograph (DEIMOS, \citealt{Faber+2003}) on 2013 June 9 in the $R$ band and also reduced using standard techniques.  Photometry was performed in IDL using a 3$\arcsec$ radius aperture relative to unsaturated SDSS stars in the field of view, with $riz$ transformed to $R$ using the Lupton transformation equations.

\subsection{RT-22}
\label{sec:rt22}

Observations of GRB 130427A  were completed with the 22~m radio telescope RT-22 at 36.8~GHz in the Crimean Astrophysical Observatory using modulation radiometers.  We applied the standard ``ON--ON'' observing technique, when antenna temperatures are recorded while pointing the two different beam lobes with orthogonal polarizations at the target in turn. We took several series containing from 30 through 36 measurements of 300~s exposure. The average temperature value and its dispersion were estimated. The orthogonal polarization of the beam lobes provides an intensity estimate irrespective of source polarization. The antenna temperature was corrected for extinction in the atmosphere, and a flux density was estimated using observations of calibration sources.  The flux was also corrected for an elevation dependence of the effective area of the radio telescope.

\subsection{CARMA}
\label{sec:carma}

We observed the position of GRB\,130427A with the Combined Array for Research in Millimeter Astronomy (CARMA) on several occasions over three days following the burst, including two separate epochs on the first day after the GRB.  All observations were conducted in single-polarization mode with the 3 mm receivers tuned to a frequency of 93 GHz, and were reduced using MIRIAD using observations of 3C~84 and 3C~273 to calibrate the flux scale. 

\subsection{PdBI}
\label{sec:pdbi}

The IRAM Plateau de Bure Interferometer (PdBI, \citealt{Guilloteau+1992}) using the Wideband Express (WideX) correlator was pointed to the GRB 130427A location on six occasions at 86.7~GHz in configurations of six and five antennas. The millimeter counterpart was detected 3 days after the GRB onset with a high ($\sim10$) S/N ratio at a distance of $\delta_{\rm RA} =-0{\farcs}52 (\pm 0{\farcs}13)$ and $\delta_{\rm Dec}=-0{\farcs}30 (\pm 0{\farcs}15)$ from the phase center coordinates.  Photometry at each epoch was performed using UV point-source fits to the phase center using the GILDAS software package.
The primary flux calibrator was MWC349. It was used on June 15 to derive
the flux of the amplitude/phase calibrator 1156+295 to 1.35 Jy at the
observing frequency. This flux was then used over the whole monitoring
period.

\subsection{VLA}
\label{sec:vla}

The position of GRB\,130427A was observed with the Karl G. Jansky Very Large Array (VLA) in its D, DnC, and C configurations, under programs 13A-411 (PI: A. Corsi), 13A-046 (PI: E. Berger), S50386 (PI: S. B. Cenko), and SE0851 (PI: A. Fruchter). The data were taken between 2013 April 27 and 2013 September 02 in several bands (K, Ka, Ku, X, C, and S; L observations were also acquired but are not presented here owing to contamination from sidelobes of a nearby object), covering overall a frequency range between 1.4\,GHz and 37\,GHz (see Table \ref{tab:radiophot}). All of the observations were conducted using the standard WIDAR correlator setting (8-bit sampling, yielding a total bandwidth of 2\,GHz per band), except a few X-band observations used in reference pointing (which employed a narrower bandwidth). 

Data were reduced using the Common Astronomy Software Applications (CASA), with the exception of a few observations in mid-May which  were reduced using the Astronomical Image Processing System (AIPS). 3C 286 was used for bandpass and flux calibration; J1125+2610 and J1159+2914 were used for gain and phase calibration. Uncertainties in the measured flux were calculated as the quadrature sum of the map root-mean square (RMS) and a fractional systematic error (of order 5\%) to account for uncertainty in the flux-density calibration.

\subsection{GCN Circulars and Other Sources}
\label{sec:gcn}

While nearly all of the optical photometry of the burst and afterglow is from our own observations, we use some data from the literature and other sources to supplement these during gaps in our temporal coverage: specifically, the early-time $R$ measurements of \cite{GCN14476} (taken during prompt emission and only used in plotting), $ugriz$ photometry of \cite{GCN14478}, $BVRI$ photometry from \cite{GCN14596}, and $JH$ photometry from \cite{GCN14483,GCN14506,GCN14514}.  The uncertainties in all of these GCN-derived points are increased to 0.05 mag in our modeling if a smaller error was quoted.  Our late-time observations of the SN are also supplemented by the $r$-band photometry of \cite{Xu+2013}, although we note that their earlier $r$-band measurements show an offset from our own P60 photometry at similar epochs.
We also take LAT observations from Figure 2 of \cite{Tam+2013} and radio observations (GMRT and a few supplementary CARMA and VLA points) from \cite{Laskar+2013}.

\subsection{Host Galaxy}
\label{sec:host}

A relatively bright, extended host galaxy is present underlying the GRB in SDSS archival images.  If not taken into account, the additional flux from this source would have significant impact on the modeling of the afterglow and SN.  While we correct for this directly in the P60 reductions via image subtraction against SDSS reference imaging, this technique is not applicable for non-SDSS filters and would be impractical for use across our entire dataset given the large number of instruments employed.

To correct for the host contribution to the afterglow, we first downloaded optical ($ugriz$) photometry of this object from the online DR9 catalog of the SDSS.  While this is adequate for constraining the contribution of the host to the $gri$ filters, the SDSS $u$ and $z$ detections are marginal, and other filters are not covered by the survey.  For other bands in which we have late-time observations ($t>20$ day) and are far from the spectral energy distribution (SED) peak of the expected supernova (specifically, the UVOT UV filters and the NIR $H$ and $K$ bands), we proceed by assuming that the late-time decay is fit by a power-law extrapolation of the earlier measurements using our model and fit a constant component to the late-time data to estimate the host contribution to these bands.

Finally, we interpolate to the remaining filters (including $u$ and $z$) by fitting a stellar population to the UV/optical/NIR photometry (using the same technique as in \citealt{Perley+2013a}) fixed at the GRB redshift and performing synthetic photometry in the remaining desired filters.  The corresponding fluxes were then subtracted from all (non-P60) photometry measurements before modeling and analysis.  The magnitudes used in this procedure are presented in Table \ref{tab:host}.

A more detailed analysis of the host galaxy is reserved for future work, although for completeness we report the basic parameters of the host as determined by our SED fit here: we find a stellar mass of $M_* = (2.1 \pm 0.7) \times 10^{9}~{\rm M}_\odot$, a mean population age of 250 Myr, and a small amount of extinction ($A_V \lesssim 0.5$ mag).  These properties --- indicating a blue, young, low-mass galaxy --- are quite typical of the low-$z$ GRB host population \citep{Savaglio+2009}.

\section{GRB and Afterglow Behavior and Rest-Frame Comparisons}
\label{sec:behavior}

\subsection{Prompt Emission Spectral Properties and Isotropic-Equivalent Energetics}
\label{sec:eiso}

GRB\,130427A is unquestionably an exceptional event.  Its bolometric fluence of $2.68$~$\times$~10$^{-3}$ erg cm$^{-2}$ is the largest of any GRB detected by the all-sky Konus satellite in almost two decades of operation \citep{GCN14487}; indeed, it is the first GRB with a fluence value exceeding 10$^{-3}$ erg cm$^{-2}$ recorded since 1988 \citep{Mitrofanov+1990,Atteia+1991}.  Only one GRB in history is known to have exceeded it, GRB 840304 \citep{Klebesadel+1984}\footnote{GRB 830801 probably also had a higher fluence, although its exact value is uncertain \citep{Kuznetsov+1987}}.

The brightest GRBs originate from events that are in relatively close proximity to Earth (as in the case of GRB 030329; e.g., \citealt{Price+2003,Hjorth+2003}) or have exceptional luminosity (as in GRBs 990123 or 080319B; e.g., \citealt{Andersen+1999,Kulkarni+1999,Bloom+2009,Racusin+2008,Wozniak+2009}).  In the case of GRB\,130427A, both factors play a role.  At the observed redshift of this GRB, the observed fluence corresponds to an isotropic-equivalent (that is, not beaming-corrected) energy release of $E_{\gamma,{\rm iso}} = 8.5 \times 10^{53}$ erg ($= 0.5~{\rm M}_{\sun}c^2$).  In Figure \ref{fig:eiso} we plot this value in comparison to a wide range of other GRBs taken from the literature:  \Swift (from \citealt{Butler+2007}, using the most up-to-date catalogs online\footnote{http://butler.lab.asu.edu/Swift/index.html}), Konus (from a search of the GCN Circulars from 2005 onward), \textit{Fermi}-GBM (from \citealt{Goldstein+2012} and \citealt{Paciesas+2012}), and a variety of pre-\Swift satellites from \cite{Amati+2006}.  Fluences were converted to $E_{\gamma,{\rm iso}}$ when the latter was not calculated explicitly. To avoid redundancy in cases where multiple satellites detected a GRB, we plot the Konus value in preference to GBM, and GBM in preference to \Swift.

\begin{figure}
\centerline{
\includegraphics[scale=0.6,angle=0]{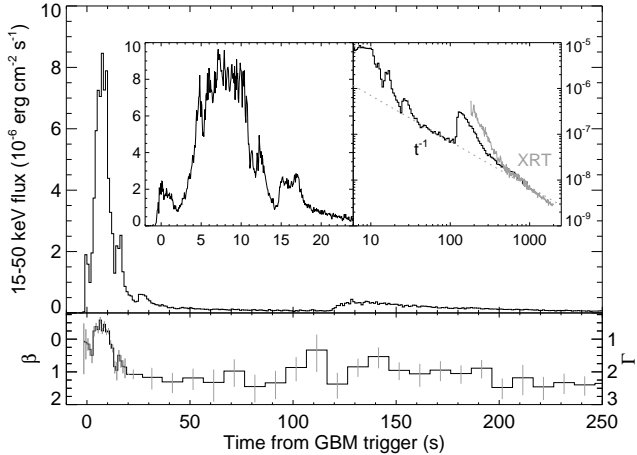}} 
\caption{Soft gamma-ray (15--50 keV observed) light curve from the \Swift BAT, showing the prompt emission and early afterglow.  The burst is dominated by a brief, extremely intense episode 0--20~s post-GBM trigger (shown binned at 64 ms in left inset).  The inset at right shows a logarithmically binned and scaled version out to 2000 s, demonstrating the slow power-law-like ($F \propto t^{-1}$) decay of the gamma-ray flux suggestive of a gamma-ray afterglow component.  A scaled version of the XRT flux is also plotted in this inset, showing similar temporal evolution except during the probable X-ray flare at 120--400 s.}
\label{fig:batlc}
\end{figure}

GRB\,130427A is comparable to the most energetic bursts in any of these populations; its $E_{\gamma,{\rm iso}}$ is 1--2 orders of magnitude above the median value for \Swift GRBs and significantly higher, even, than the average for GRBs detected by Konus (which is insensitive to the faintest events).   It would be easily detectable at almost any redshift; events with one tenth this $E_{\gamma,{\rm iso}}$ routinely trigger \Swift out to $z>5$ (and even $z>8$).   Nevertheless, GRB\,130427A is not unprecedented by GRB standards; several dozen known events outrank it in $E_{\gamma,{\rm iso}}$.   GRB\,130427A is remarkable because it is by far the closest event with a large energy release:  all previous events of similar or greater energetics have been at $z>0.9$ (corresponding to a factor of 10 in $d_L^2$); the next-most-luminous event to occur comparably close (at $z<0.5$) was GRB 030329, which was a full factor of 50 lower in $E_{\gamma,{\rm iso}}$.

In brief, in terms of apparent energetics, GRB\,130427A constitutes a highly but not exceptionally luminous GRB, seen closer than any burst of comparable luminosity in the afterglow era.  It therefore represents the best chance to date to study the properties of a high-luminosity GRB in the extreme detail afforded by such nearby events.

\subsection{Prompt-Emission Temporal Properties}
\label{sec:batlc}

\begin{figure}
\centerline{
\includegraphics[scale=0.55,angle=0]{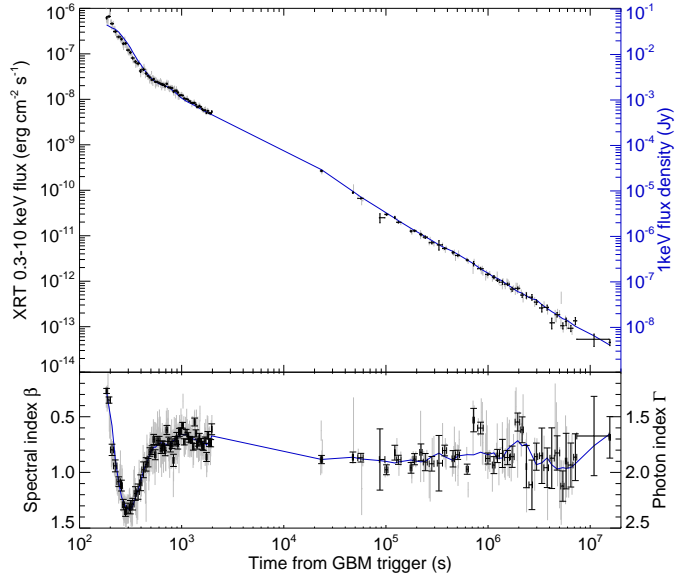}} 
\caption{X-ray (0.3--10 keV observed) light curve from the \Swift XRT.  Gray shows moderately binned XRT observations as taken from the \Swift Burst Analyser; black points are more significantly binned.  The blue line shows the effective 1 keV flux density calculated given the total flux and spectral index, smoothed over nearby data points (the blue line in the lower plot shows the smoothed value of the spectral index used in this procedure).  Except at early times during the final prompt-emission episode, the afterglow shows only minor deviations from a single power law ($\alpha = 1.35$) throughout.  In particular, there is no evidence of a jet break out to at least 100 days. }
\label{fig:xrtlc}
\end{figure}

The hard X-ray temporal profile of the GRB (15--50 keV) from the BAT is plotted in Figure \ref{fig:batlc}.   While the burst appears very long in the BAT band ($T_{90}$ = 163 s, with detectable emission continuing to $\sim 2000$~s; \citealt{GCN14470}), this is largely because of the presence of a long-lived, shallow power-law component extending throughout the initial period of observations following the burst itself and a rebrightening episode beginning at $t=120$ s.  Beyond these two features, the burst is dominated by an initial pulse complex from $-1$~s to 18~s that contributes about 60\% of the fluence in BAT's softer channels and the large majority of the fluence in the true gamma-ray bands; for example, the Konus light curves of \citet{GCN14487} suggest that in the 300--1160 keV band encompassing $E_{\rm pk}$, over 95\% of the total fluence was emitted between 4 s and 11 s.

The rebrightening episode from 120 s to 350 s was caught starting at 195 s by the \Swift XRT and shows some interesting features in that band.  While the overall peak energy $E_{\rm pk} \approx 240$ keV reported by Konus is well inside the gamma-ray band, the XRT spectral index (Fig. \ref{fig:xrtlc}) evolves strongly during the observation and is softer than $\beta>1.0$ at the end of the flare ($t > 220$ s), indicating a peak below $\sim 1$~keV at that time.  Only minimal evolution is seen in BAT and the spectral index is never particularly hard even at the start of the flare, which may indicate unusual spectral structure.  As this paper does not focus on the details of the prompt emission, we leave further study for future work.  Its most pertinent characteristics for our analysis of the afterglow is that it is clearly associated with the prompt emission (on the basis of its peaked spectral profile and hard-to-soft evolution), is energetically subdominant compared to the primary pulse, and (as discussed in the next paragraph) appears to have only minimal effect on the underlying afterglow light curve.

The nature of the long-lived shallow power-law component, however, is of significant interest for our subsequent analysis.  The appearance of such a component in a $\gamma$-ray light curve is unusual; to our knowledge the only previous \Swift GRBs showing a similar signature is GRB 080319B, although similar features have been searched for in bright BATSE bursts \citep{Giblin+2002} and probably seen in at least one case \citep{Giblin+1999,Fraija+2012}.  A logarithmically binned BAT light curve is plotted in the inset to Figure \ref{fig:batlc} along with a rescaled XRT light curve; the dashed line shows a power law fading as $F \propto t^{-1}$ which (except in the region of the flare discussed in the previous paragraph) matches both light curves reasonably well both before and after the late-time prompt emission episode.  The photon index is close to $\Gamma=2$ (spectral index $\beta=1$) in both the BAT and XRT bands, as is the spectral index connecting the two bands.  These properties --- namely, a power-law SED with $\beta \approx 1$ and power-law light curve with a shallow decay index --- are highly suggestive of afterglow emission associated with the forward shock.  Indeed, our later modeling (\S \ref{sec:modeling}) strongly supports this hypothesis as the existence of emission in the BAT band with flux, temporal evolution, and spectral index similar to what is observed is inevitably predicted based on simple extrapolation of the X-ray flux in time and frequency.

\subsection{X-ray Afterglow Behavior and Luminosity}
\label{sec:xraylc}

The X-ray light curve (Fig. \ref{fig:xrtlc}) shows very simple evolution.  After the end of the prompt emission episode at $\sim 400$~s, the X-ray flux fades as an effectively unbroken power law (a small amount of excess is seen during a short observation at $t=0.27$ days) for the entire remainder of the observation, a span of 5 orders of magnitude in time.  In particular, the light curve does not show any sign of steepening at late times, indicating a very late jet break ($t > 100$ days), placing a limit on the collimation angle, and giving a lower bound on the true energy scale (see \S \ref{sec:energetics}).

Like the GRB itself (and like the optical afterglow, Fig.\ \ref{fig:optlc}), the X-ray afterglow is remarkably bright.
In the top panels of Figure \ref{fig:comparexo} we plot the \Swift XRT light curve against a sample of other XRT light curves (limited for clarity to subsamples of the closest events, the most energetic events, and an effectively random subsample of all \Swift GRBs).  Except between 0.02 and 0.4 days (when it is exceeded by GRB 111209A), GRB\,130427A is the brightest X-ray afterglow to be observed by the satellite at any common time of comparison.  (By comparison to \citealt{Jakobsson+2004}, it is also brighter than any pre-\Swift X-ray afterglow.) In an absolute sense, however, its properties are much less exceptional:  its X-ray luminosity is only slightly above average for \Swift GRBs.

\subsection{Optical Afterglow Behavior and Luminosity}
\label{sec:optlc}

\begin{figure*}
\centerline{
\includegraphics[scale=0.7,angle=0]{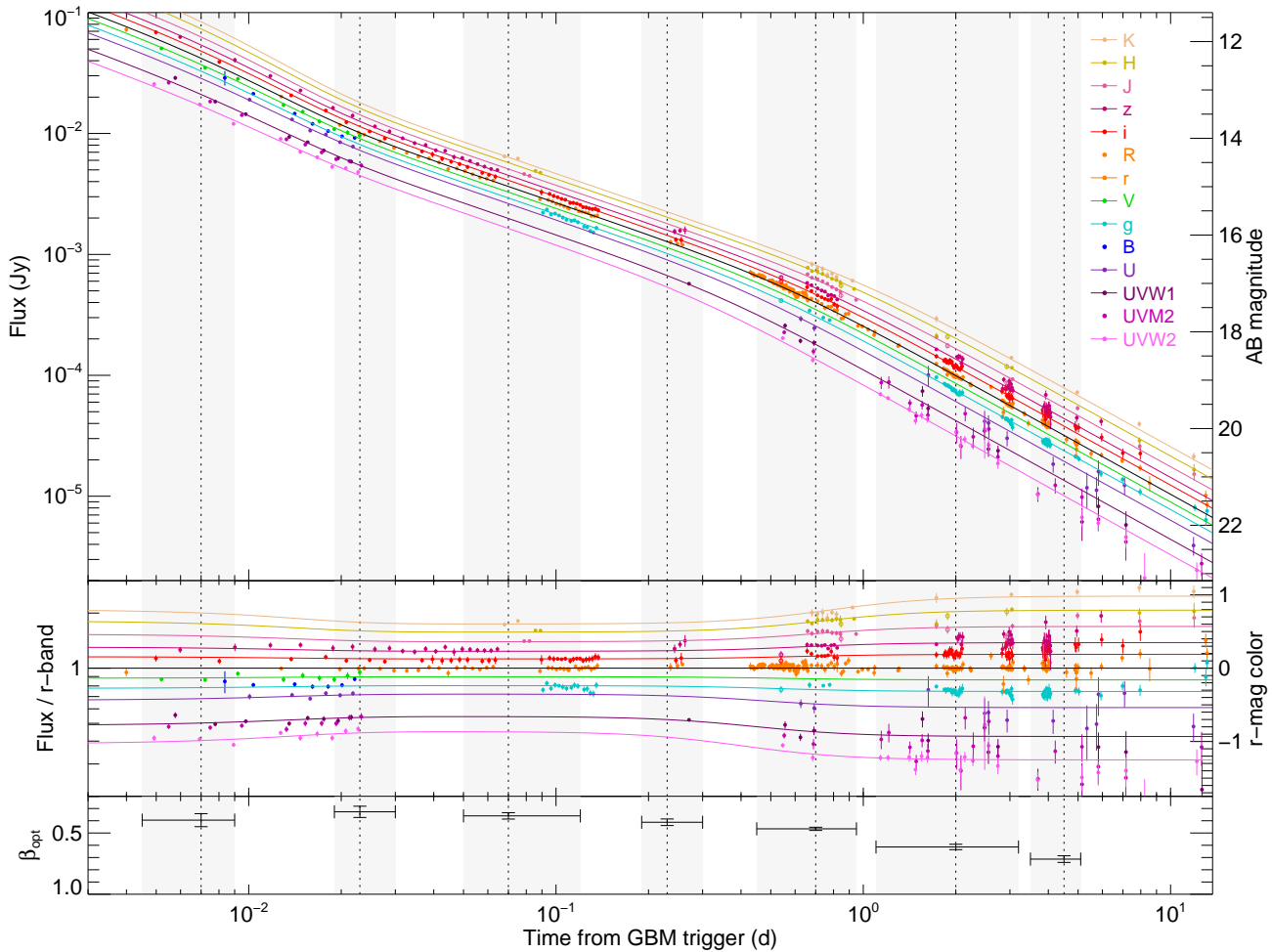}} 
\caption{UV/optical/NIR light curves (various filters spanning 180--2200 nm) from the \Swift UVOT and our ground-based campaign, supplemented by a few points from the GCN Circulars.  Points are color coded by filter and magnitudes are not corrected for host or Galactic extinction.  The $UVM2$, $B$, and $R$-band points have been shifted slightly to match other, nearby filters.  All points were fit by an empirical model including color change; clear blue-to-red evolution is seen over the latter half of the light curve, with probable red-to-blue color evolution in the early phase.  The top panel shows absolute fluxes; the middle panel shows colors relative to the (modeled) $r-$band curve.  We also created a series of coeval SEDs by epoch (grey bands) and fit for the spectral index $\beta_{rm opt}$ in each segment directly (bottom panel).}
\label{fig:optlc}
\end{figure*}

The wide-field RAPTOR monitored the evolution of the optical afterglow throughout the gamma-ray activity, recording a peak of $R \approx 7$ mag \citep{GCN14476}, which would make this the second-brightest (at peak) optical afterglow yet observed, brighter than GRB 990123 \citep{Akerlof+1999} but not as bright as GRB 080319B, the ``naked eye'' burst \citep{Racusin+2008}.   Our own observations do not cover this prompt phase, but begin at 315~s after the GBM trigger and provide nearly continuous coverage (in the $r$/$R$-band, and more partial coverage in other bands) for the next two days, followed by nightly coverage for the next several weeks through the emergence and peak of the associated supernova.

A plot of the flux of the optical afterglow in various filters as a function of time is shown in Figure \ref{fig:optlc}.  Generically, the optical afterglow evolution during the course of our observations can be described as three power-law segments: a moderately rapid decay at early times ($<0.02$ d; decay index $\alpha=1.15$), a more gradual decay at intermediate times (0.02--0.6 day; $\alpha=0.85$), and then another moderately rapid decay at late times ($>0.6$ days; $\alpha = 1.45$).  Beyond about 6 days the optical afterglow flux is significantly contaminated by host and SN emission; while our observations are consistent with an unbroken decay (particularly in the NIR bands, where the SN and host are relatively faint), we will ultimately require late-time reference images to confirm this unambiguously.

We fit the optical afterglow observations (for points at $t<6$ days; later points are not used owing to uncertainty about the SN and host contributions) in all filters simultaneously using the empirical model of \cite{Perley+2008b}, which treats the light curve as a sum of \cite{Beuermann+1999} smoothly broken power laws; color evolution is incorporated in this model as the result of different intrinsic spectral indices underlying each component as well as the rising and falling power-law segments of each component.\footnote{We note for clarity that this \emph{empirical} model, which is used here to qualitatively understand the behavior of the optical afterglow and to assist in the interpolation of the observations to a common epoch, is not the same as the \emph{physical} model we will use in \S \ref{sec:modeling} to interpret the entire multiwavelength dataset.}  Only two components are required to reproduce the optical behavior without introducing any significant residual trends: a relatively fast-decaying component at early times plus a more gradually evolving component (with a break at $t=0.7$ d) to explain the remainder of the evolution.  The early- and late-time colors are similar, but the shallow-decay period at 0.1--0.7 day shows a significantly flatter (bluer) spectral index (although the difference is relatively small; $\Delta\beta = 0.34$).  This color evolution can be seen more clearly in the middle panel of Figure \ref{fig:optlc}.

To examine the temporal evolution of the optical SED of this GRB in more detail and verify that the color evolution indicated by our model fits is real, we subdivided the afterglow into seven temporal ranges:  $t=$ 0.0045--0.009 d, 0.019--0.03 d, 0.05--0.12 d, 0.19--0.3 d, 0.45--0.95 d, 1.1--3.2 d, and 3.5--5.1 d.  We then performed separate fits to the light curve on these epochs individually to create an optical SED at approximately the midpoint of each of these these epochs.  The interpolated SEDs, corrected for Galactic extinction ($E_{B-V} = 0.02$ mag; \citealt{SFD,Schlafly+2011}\footnote{\cite{Peek+2013} have recently found evidence for nonstandard extinction properties at high Galactic latitudes in the UV, so the exact correction has some additional uncertainty outside the optical bandpasses.  Fortunately, the absolute value of the extinction is low and the impact of this effect should be relatively minor.}), are presented in Table \ref{tab:seds}.

It is also desirable to constrain and correct for the effects of host extinction.  To do so, we take the SED at $t=2$ days (which is after the temporal break and should be minimally affected by intrinsic curvature in the SED resulting from a break passage or contribution from the early-time component) and fit these observations assuming an intrinsic power law with spectral index $\beta_{\rm O}$ and a \cite{Fitzpatrick1999} extinction model with a profile similar to that of the Large Magellanic Cloud \citep[LMC;][]{Misselt+1999} and $R_V=3$. (The extinction column toward this GRB is too low to robustly constrain the choice of actual extinction law, although models with a 2175~\AA\ bump marginally better reproduce the SED than the featureless SMC-like curve.)  We measure a small but significant amount of extinction ($A_V = 0.13 \pm 0.06$ mag) from this fit.  We then fit to all seven epochs for the spectral slope $\beta_{\rm O}$ after correcting for host extinction using this value.  The result (see the bottom panel of Fig.~\ref{fig:optlc}) confirms the blue-to-red evolution between moderately early ($t = 10^{-2} - 10^{-1}$ d) and late ($t > 1$ d) times.  The red-to-blue transition from the earliest observations during the initial steep decay to $t>10^{-2}$ s is also seen, although it is not highly significant as we do not have NIR observations during this early period.

\begin{figure*}
\centerline{
\includegraphics[scale=0.65,angle=0]{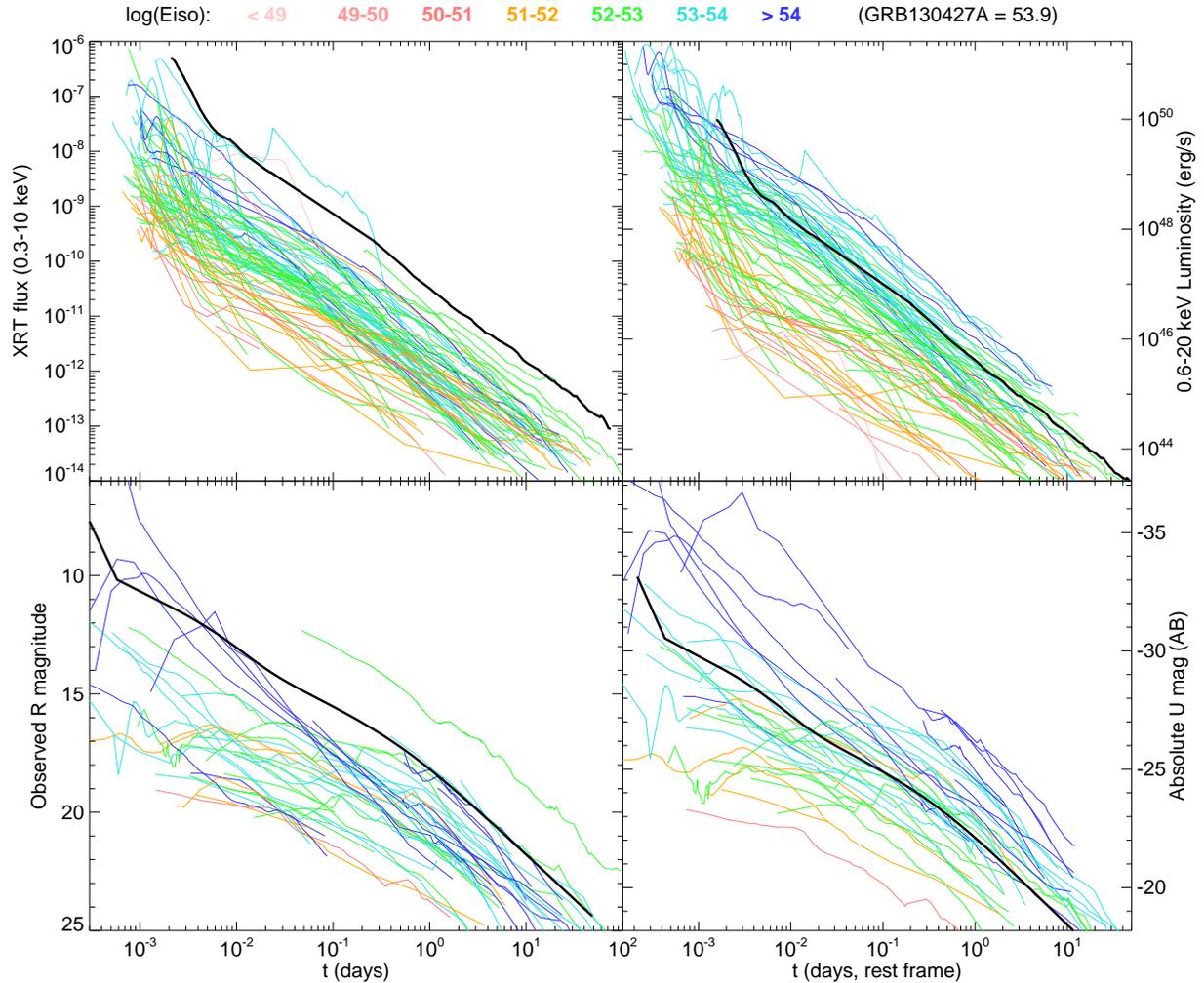}} 
\caption{Light curves of GRB\,130427A in the X-ray and optical bands, compared against literature samples of other GRBs, both in the observer frame (left panels) and shifted to a common redshift (right panels).   All curves are color coded by $E_{\gamma,{\rm iso}}$, with less luminous GRBs in red and orange, moderately luminous GRBs in green, and very luminous GRBs in cyan and blue.  GRB\,130427A (which would be at the upper end of the cyan range) is plotted as a thick black line.  Comparison samples for the X-ray curves are taken from the \Swift repository for GRBs at $z<0.8$ plus Konus-detected \Swift GRBs (we also include GRB 030329; \citealt{Tiengo+2003,Tiengo+2004}); optical comparison samples are taken from a subsample of events from the database of \cite{Kann+2006,Kann+2010,Kann+2011} (specially, all events at $z<1$, with $E_{\rm,iso} > 10^{53}$ erg and $1<z<2$, or with $E>10^{54}$ erg at any redshift) and from \cite{Cenko+2009}.  These cuts are chosen to reduce the number of comparison light curves to a manageable number and limit the sample to bright bursts with well-determined $E_{\rm, iso}$ values and high likelihood of optical afterglow follow-up, while still sampling the entire range of prompt-emission luminosities.)
}
\label{fig:comparexo}
\end{figure*}

While the optical afterglow of GRB\,130427A is the second-brightest (in terms of apparent magnitude, compared at a common time-post-GRB) of any GRB observed to date for the large majority of its evolution (GRB 030329 was brighter by about 1.0 mag), as at other wavelengths this apparent brightness relative to other GRBs is primarily a function of the burst's proximity: as shown in Figure \ref{fig:comparexo}, its overall luminosity and decay rate are quite characteristic of other optical afterglows observed for moderately luminous bursts.

\begin{figure*}
\centerline{
\includegraphics[scale=0.65,angle=0]{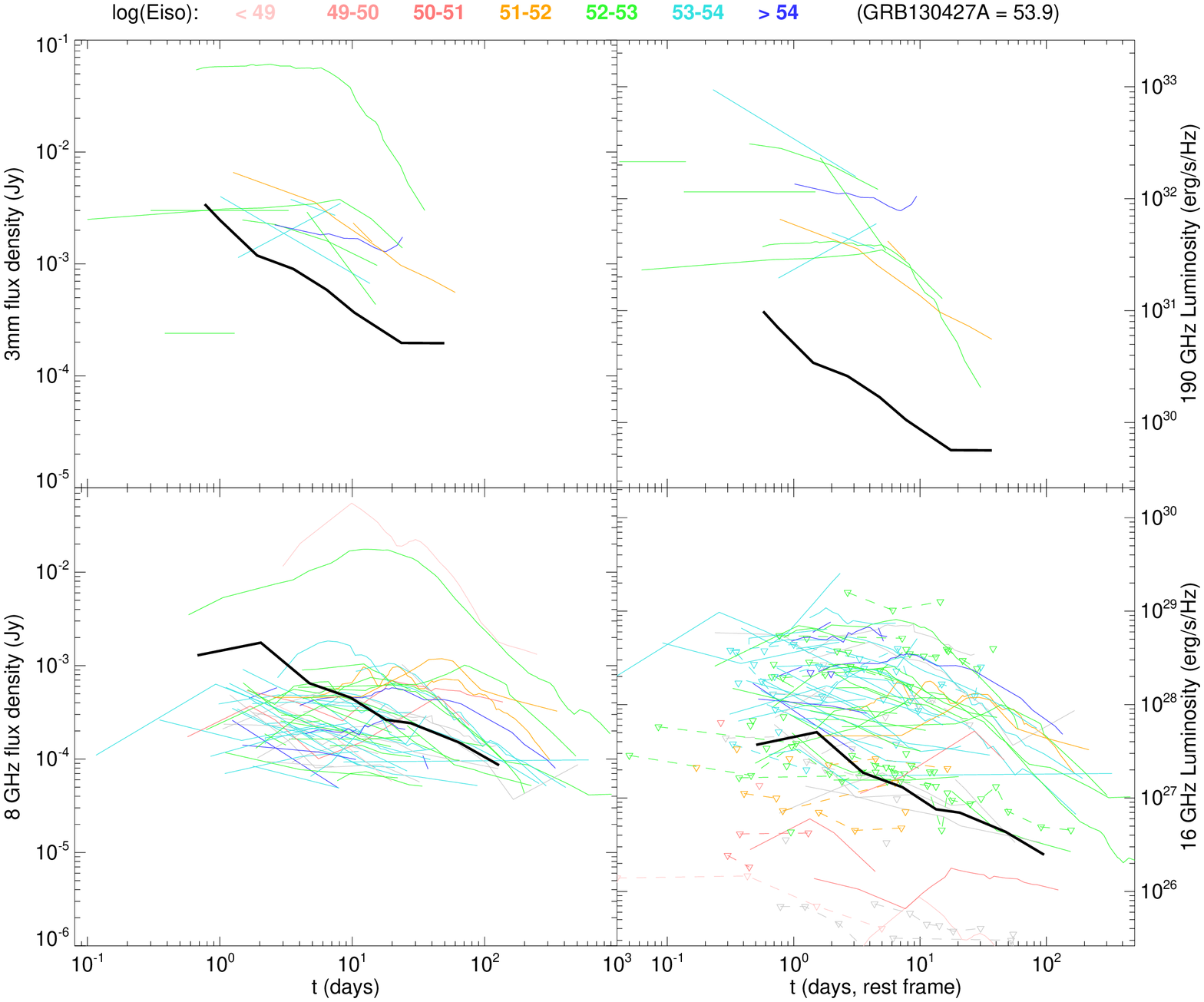}} 
\caption{Light curves of GRB\,130427A at millimeter and radio wavebands compared with literature samples of other GRBs, both in the observer frame (left panels) and shifted to a common redshift (right panels).   Curves are color coded by $E_{\gamma,{\rm iso}}$ as in Figure \ref{fig:comparexo}.  Comparison samples for the millimeter light curves are taken from \cite{deUgarte+2012} and work in preparation by Perley et al.\ and Laskar et al.; radio light curves are from \cite{Chandra+2012}.  The radio afterglow of GRB\,130427A is actually underluminous relative to most \emph{radio-detected} GRBs, but as a significant fraction (about half) of GRBs are not detected at radio wavelengths it may not be intrinsically unusual in this regard.}
\label{fig:comparemr}
\end{figure*}

\subsection{Radio Afterglow Behavior and Luminosity}
\label{sec:radiolc}

While we were not able to begin observing at radio wavelengths for several hours following the burst and cannot directly constrain the early-time behavior, GRB\,130427A nevertheless represents one of the most thoroughly observed radio afterglows to date, with observations spanning $< 1$--95 GHz in frequency and 0.36--128 days in time.  Interestingly, the properties of the afterglow at these wavelengths (luminosity, spectral shape, and temporal behavior) are quite \emph{unlike} those of most other GRB afterglows detected at radio wavelengths thus far.

First, the GRB\,130427A afterglow is actually remarkably \emph{faint} given the close proximity and large energetics of the GRB (Fig. \ref{fig:comparemr}).  For example, its millimeter afterglow is a factor of 100 less luminous than that of GRB 030329 at $t=3$ days.  At $t=10$ days the radio afterglow is $\sim10$ times less luminous than that of GRB 030329 at the same epoch and a factor of 100 below the most luminous late-time radio GRBs.   While lower-luminosity afterglows are not unprecedented, nearly all of them correspond to very nearby events that were also underluminous in gamma-rays (and indeed at all other wavelengths).

The light curve is also remarkable.  The early CARMA observations show rapid fading beginning at $t=0.7$ days ($\alpha \approx 1.4$).  This rapid evolution slows down at later times:  the later CARMA and subsequent PdBI points are consistent with approximately $\alpha \approx 0.4$, and the high-frequency VLA bands show slow similar temporal evolution.  In contrast, most well-studied radio afterglows are flat or even rising over the first week and fade steeply after that time \citep{Chandra+2012}.

Finally, the SED is unusual: typically the early-time radio spectral index of a GRB is relatively hard either because of synchrotron self-absorption (which predicts $\beta \approx -2$ below the self-absorption frequency) or because the observations are below the synchrotron peak (in which case $\beta \approx -1/3$).  We do not see any sign of self-absorption here except possibly during the first epoch, when the spectral index as measured by a comparison of the two C-band intermediate frequencies (5.1 and 6.8 GHz, acquired simultaneously) is remarkably hard and consistent with $\beta \approx -2$.  At all other times the radio spectral index is very close to flat ($\beta \approx 0$) with no sign of a turnover toward low frequencies.

Given that the GRB is ``normal'' in nearly all other ways, it would seem like a curious coincidence that its radio properties are so unusual.  However, it is important to recognize that in the radio and millimeter bands, the comparison sample of GRBs is highly flux-limited: only about 50\% of all GRBs are detected at radio wavelengths \citep{Chandra+2012}.\footnote{This is not the case in the X-ray or optical bands, where almost all bursts detected in gamma-rays are also detectable at these wavelengths if observations are executed promptly with a $>1.5$~m telescope \citepeg{Cenko+2009}.  A sole exception is the category of optically dark GRBs (which constitute about 15\% of the population; \citealt{Perley+2009,Greiner+2011}; the nondetection of these events is primarily due to dust extinction, an extrinsic factor.)}  Indeed, GRB\,130427A would likely not have been detected itself at radio wavelengths prior to the VLA's upgrade had it occurred at a ``typical'' \Swift redshift of $z>1.5$.   It is therefore more likely that this otherwise entirely ordinary GRB is not particularly exceptional or unusual at radio wavelengths, but instead that we have obtained our first clear look at a member of the (large, but previously poorly understood) radio-faint population by virtue of this event's close proximity.  We will return to the discussion and interpretation of the radio behavior of this GRB afterglow in \S \ref{sec:fwdshock}.

Radio observations of point-like sources can be affected by interstellar scintillation.  We do not see any obvious effects of scintillation in our observations (all well-sampled radio light curves and SEDs show a relatively smooth appearance, except perhaps for the first C-band epoch), but scintillation nevertheless could represent a significant additional source of uncertainty if present.   To estimate the possible scintillation contribution, we use the angular-diameter version of Equation 2 in \cite{Taylor+2004} (using our inferred values of $E_K$ and $A_*$ from \S \ref{sec:modeling}) to estimate the size of the GRB as a function of time relative to the angular scintillation scale in Figure 2 of \cite{Walker2001}.  The point-source scintillation scale at this Galactic latitude of 4.5 $\mu$as is much smaller than the expected angular size of $\sim 40\, t_{\rm day}^{0.75}$~$\mu$as, indicating that any scintillation should be strongly damped at high frequencies ($\nu \gtrsim \nu_o$; the transition frequency $\nu_o$ is $\sim 5$~GHz at this Galactic latitude) and effectively negligible.  The angular scale, however, is larger at lower frequencies where the strong scattering regime applies, scaling as $\theta \propto (\nu/\nu_o)^{-11/5}$ \citep{Walker1998}.  At the time of the first low-frequency observations ($\sim 5$~days), the scintillation scale at 1.5~GHz is $\sim 100$~$\mu$arcsec, which is similar to the source size at that time, and modulation of up to $(\nu/\nu_o)^{17/30}$ or about 40\% can be expected, which should be considered when interpreting the GMRT and L/S-band observations.

\subsection{Supernova Behavior and Luminosity}
\label{sec:snlc}

The appearance and fading of a SN is unmistakable in the late-time data.  Some caution must be used in interpreting these observations in more than a qualitative sense, since large systematic uncertainties are introduced at late times owing to the contribution of the host and afterglow, which are known approximately but have significant uncertainty.

\begin{figure*}[ht!]
\centerline{
\includegraphics[scale=0.64,angle=0]{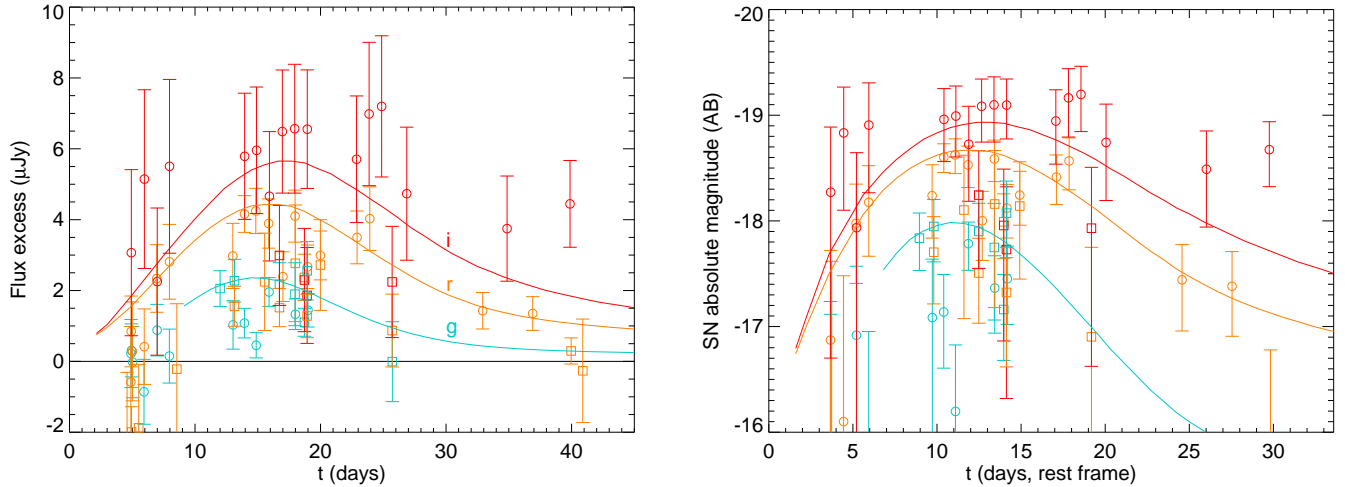}} 
\caption{Host- and afterglow-subtracted light curves of SN 2013cq associated with GRB\,130427A.  The left curve is linear in time and flux space and shows the rise and fall of the SN, although uncertainties are significant owing to the uncertain host contribution.  The right curve shows the absolute magnitude of the associated SN relative to scaled templates of 1998bw; we find that a slightly less luminous and faster-timescale version of SN 1998bw reasonably matches the data.  Circles indicate P60 observations with the host subtracted in image space, square symbols indicate standard aperture photometry with the host in flux space.}
\label{fig:sncompare}
\end{figure*}

Nevertheless, the current observations are sufficient for a preliminary investigation into the basic properties of the associated SN.  In Figure \ref{fig:sncompare}(a) we plot the excess flux after subtraction of the power-law afterglow (based on our preferred model fit in \S \ref{sec:optlc}, which only uses data from the first 5 days after explosion to minimize any SN contribution), which shows the clear rise and fall of an additional, red component.  The $i$-band signature currently has large systematic errors because of the uncertain host contribution in this band which we expect will be significantly reduced after late-time reference imaging is available; for now we do not include these bands in the SN fit.  The $g$ and $r$-band observations are fit with a SN 1998bw template taken from an interpolation of the observations of \cite{Galama+1998} and \cite{Clocchiatti+2011}, scaled independently by linear factors in time and luminosity.  We find that a SN with an apparent luminosity (\emph{not} corrected for extinction) $\sim 0.6\times$ that of SN 1998bw and a timescale 0.8$\times$ that of SN 1998bw provides a reasonable fit to our observations, in good agreement with the analysis of \cite{Xu+2013}.  After correction for host extinction based on our estimate of $A_V = 0.13$ mag (\S \ref{sec:optlc}) the intrinsic luminosity is 0.7$\times$ that of SN 1998bw.

A more detailed examination of the SN properties will be left for further work and will not be discussed here.  We did verify that the SN should not contribute a large amount of flux to the UV or optical filters that might have affected our host-galaxy subtraction in \S \ref{sec:host}; assuming colors similar to those of \cite{Simon+2010} and \cite{Kocevski+2007}, the contribution to these bands is relatively small ($\lesssim$15\% of the host+afterglow flux) at all times.

\section{Multiwavelength Modeling and Interpretation:  A Classic Case of Reverse and Forward Shocks}
\label{sec:modeling}

\subsection{Construction of Coeval SEDs}

Our multifrequency, multi-epoch data provide a powerful test of basic afterglow models.  The optical, X-ray, and radio fluxes and spectral slopes are well-observed and independently constrained at almost every epoch, permitting us to construct an evolving SED spanning many orders of magnitude in both time and frequency.

To provide the best constraints possible and minimize the uncertainties associated with the nonsimultaneity of observations in each band for this exercise, it is necessary to interpolate measurements to construct coeval SEDs at a series of semiregularly spaced epochs spanning the full range of observations.  Optical observations have already been interpolated in this way using the technique described in \S \ref{sec:optlc}; here we extract the XRT, millimeter, and multi-frequency radio fluxes (as well as BAT observations during the first epoch) and XRT spectral index at each of these same epochs (0.007, 0.023, 0.07, 0.23, 0.7, 2.0, and 4.5 days), plus at a series of later-time epochs (10, 30, 60, and 130 days) after the optical light becomes dominated by host and SN.  

The BAT and XRT bands are simply directly interpolated in flux and in $\beta$ using the curves in \S \ref{sec:batxrt}.  We also interpolate (logarithmically in time and flux) the LAT data of \cite{Tam+2013} in the 0.2--10 and 10--100 GeV bins.  The radio observations are relatively sparsely sampled and the radio behavior at these wavelengths is complex and frequency-dependent, so the choice of interpolation procedure is not straightforward: fortunately, since our epoch times were deliberately chosen to be very close to the times of radio observations, the results are largely independent of the procedure.  Nevertheless, using the observed behavior of the light curves at both low and high frequencies, we choose temporal indices that provide a reasonable approximation of the observed behavior at all times to perform the interpolation as accurately as possible.  Specifically, at high frequencies ($>$50 GHz) we interpolate assuming $t^{-1.4}$ behavior at $t<1$~d, $t^{-1.0}$ at $t=1-3$~d, and $t^{-0.4}$ at later times.  At low frequencies ($<$50 GHz) we interpolate assuming $t^{-0.4}$ at $t<1$d, $t^{-1}$ between $t=1-4$~d, $t^{-0.4}$ between $t=4-100$d, and $t^{-1}$ after 100d.   Results, color-coded by epoch over the full range from 0.007--130 days, are shown in Figure \ref{fig:multised}.

\subsection{A Two-Component SED}

We first attempted to fit these SEDs with a simple single-component synchrotron SED using the standard model of \cite{Sari+1998}: a thrice-broken power law in frequency with breaks at the synchrotron self-absorption frequency $\nu_a$, the minimum electron energy $\nu_m$, and the cooling frequency $\nu_c$; spectral indices ($\beta$, where $F_\nu \propto \nu^{-\beta}$) between the breaks are $-$2, $-$1/3, \{1/2 or ($p-1$)/2\}, and $(p/2)$, where $p$ is the electron power-law index (typically $p \approx 2$--2.5) and the slope of the third segment depends on whether the electrons are fast-cooling ($\nu_c > \nu_m$) or slow-cooling ($\nu_m < \nu_c$).  Breaks are softened by means of the \cite{Beuermann+1999} function.  

Even in a qualitative sense, this spectral shape can provide a reasonable fit to the radio-optical-X-ray data \emph{only} at or after $t \approx 30$ days.  At earlier times, the flat or soft radio spectral index cannot be made to match the other bands as it would require $\nu_m \lesssim \nu_{\rm radio}$, which is not consistent with extrapolation from the optical band (it also has some difficulty explaining the shallow slope between the X-ray and optical bands $\beta_{\rm OX}$ at the earliest times).   We therefore added a second synchrotron SED to the model with the $\nu_m$ peak at low frequencies.  This two-component model was found, in general terms, to be capable of providing an excellent description of the data at every epoch.

The need for two components in this fashion is highly suggestive of the presence of both reverse and forward shocks simultaneously contributing to the afterglow, the reverse shock dominating at lower frequencies and the forward shock at higher frequencies.  The case for this interpretation grows even stronger when the temporal behavior is also considered:  the steep-shallow-steep evolution of the optical light curve and steep-shallow evolution of the millimeter light curve are very similar to the behavior of previous GRB afterglows with claimed early-time reverse shocks at these frequencies \citepeg{Akerlof+1999,Kulkarni+1999,Kobayashi+2003a,Li+2003,Wei2003,Shao+2005,Perley+2008a,Gomboc+2008,Racusin+2008,Bloom+2009,Pandey+2010,Gruber+2011,Cucchiara+2011,Zheng+2012,Gendre+2012}. In the remainder of the discussion we therefore refer to the low-frequency and high-frequency synchrotron components of the model as the reverse-shock and forward-shock components, respectively.

\subsection{Physical Modeling}

With this paradigm in mind, we attempted to reproduce the temporal \emph{evolution} of the multiwavelength SED within the standard forward/reverse-shock fireball model of GRB afterglows \citepeg{Meszaros+1993,Sari+1998,Piran+1999}.  In general, we started with attempting to reproduce the SED at $t = 2$ days post-trigger and extrapolated the SED forward and backward in time under various assumptions about the evolution of the key synchrotron parameters ($F_{\nu,{\rm max}}$, $\nu_a$, $\nu_m$, and $\nu_c$) for both the forward and reverse shocks, adjusting the assumptions as necessary to reproduce the earlier and later SEDs.  We determine the values of these parameters and do not yet attempt to solve for the underlying fundamental burst/microphysical parameters (with the exception of the electron index $p$).  Our procedure parallels the approach taken by \cite{Laskar+2013}, but was developed largely independently and is applied here to a much larger dataset, including radio observations out to $t=128$ days, significantly improved optical observations, and consideration of the spectral indices as well as the absolute fluxes.

The ingredients, assumptions, and basic results of the model are described in the following sections:

\subsubsection{Forward Shock}

The synchrotron peak frequency of the forward shock evolves as $\nu_{m,{\rm FS}} \propto t^{-3/2}$.  This is a robust prediction of all synchrotron models as it is independent of the nature of the circumburst environment and assumes only that the amount of energy in the shock remains fixed (adiabatic evolution).

The evolution of the \emph{flux} at this frequency ($F_{\nu,{\rm max, FS}}$) depends sensitively on the density structure of the circumburst medium.  
We infer a scaling very close to $F_{\nu,{\rm max, FS}} \propto t^{-1/2}$, corresponding to a wind-like density structure ($\rho \propto r^{-2}$).  Other types of profiles are strongly disfavored; in particular, the model for a constant-density environment (in which $F_{\nu,{\rm max, FS}} \propto$ const; the peak simply shifts to lower frequency) is in direct conflict with several observational aspects of the burst, in particular the continuous fading of the radio flux at 10--30 days.

Normally, the cooling frequency $\nu_c$ should increase with time in a wind-like environment ($\nu_{c,{\rm FS}} \propto t^{0.5}$).  However, such a rapid increase is not in agreement with observations:  the relatively hard X-ray slope of $\beta_X \approx 0.7$--0.9 can only be produced if the cooling break is located in or near the XRT band for the entire duration of observations.  To reproduce the X-ray evolution, we experimented with different power-law evolutions for $\nu_{c,{\rm FS}}$ with time and found a range between approximately $\nu_{c,{\rm FS}} \propto t^0$ and $\nu_{c,{\rm FS}} \propto t^{+0.2}$ to be reasonably consistent with the lack of X-ray spectral evolution; we use $\nu_{c,{\rm FS}} \propto t^{+0.1}$. 
Spectral (non-)evolution consistent with these observations could be produced by a somewhat modified circumburst density profile intermediate between the constant-density and wind-like cases (e.g., $\rho \propto r^{-1.5}$); however, we found that even small deviations away from an $r^{-2}$ profile greatly degraded the match between model and observations of the light curve.  We conclude that either the noncanonical $\nu_c$ evolution is real (which would indicate that the timescale for cooling of the fast electrons could evolve differently from what is usually assumed, perhaps owing to inhomogeneous magnetic field structure), or that some other physical process is modifying the shape of the X-ray SED.  However, since our remaining observations are below the cooling break at all times and the hydrodynamics of an afterglow in the adiabatic phase are not affected by its cooling properties, we do not consider this a significant problem for our overall model.

The self-absorption break $\nu_a$ of the forward-shock component is not seen at any point during the observations, presumably because it is below our observed frequency range; as it does not affect our results, it is set to an arbitrary low value in the modeling.  (Physical constraints later restrict it to a relatively narrow range that is consistent with its nondetection; \S \ref{sec:fwdshock}.)

The electron index $p$ is set to 2.14.  This is constrained fairly rigidly by the optical-to-X-ray spectral index, by the radio-to-X-ray spectral index at late times, and by the X-ray slope, and even relatively minor deviations from this value (greater than $\Delta p \approx 0.03$) create discernible discrepancies with the data.

Relatively soft spectral breaks are needed to avoid producing dramatic color change or spectral curvature (which are not observed) during break passages.  We used Beuermann sharpness parameters of 0.7 (near $\nu_m$) and 1.0 (near $\nu_c$).  

\subsubsection{Reverse Shock}

Predicting the power-law evolution of the reverse-shock parameters is not as straightforward as for the forward shock, since it depends on the thickness of the burst ejecta and its density profile, which in turn are produced by a combination of internal and external factors.  Basic predictions for reverse-shock emission in ejecta interacting with a wind-stratified medium were proposed by \cite{ChevalierLi2000} and \cite{Kobayashi+2003}, and further expanded upon by \cite{Zou+2005} who considered the impacts of different density profiles of the ejecta. (This model has recently been further developed and revised by \citealt{Harrison+2013}, although we do not consider these additions here.)

Two general cases are thought to apply for reverse shocks, depending on the timescale for deceleration of the ejecta (and production of the shock) relative to the timescale for the shock to propagate across the shell $t_X$. 
In the \emph{thin-shell} or Newtonian case ($t_{\rm GRB} < t_{\rm dec}$), deceleration occurs after the burst has ended and the shock remains subrelativistic, and the crossing time is set by the thickness of the shell, which in turn is set by the burst duration $t_X \approx t_{\rm GRB}$.   In the \emph{thick-shell} or relativistic case ($t_{\rm GRB} > t_{\rm dec}$), the reverse shock forms while the burst is still actively providing new ejecta for it to propagate into, allowing the shock to become relativistic during its crossing time.

In the thin-shell model, predictions for the evolution of the synchrotron parameters (after the shock peak, which is the only case we consider here since the afterglow is fading at all observed frequencies and times) depend on the power-law index of the ejecta density profile ($g$) and are given by  $\nu_{a, {\rm RS}} \propto \{ t^{-(33g+36)/(70g+35)}$, $t^{-((15g+24)p + 32g + 40)/( (14g+7)p + 56g + 28)} \} $ (for $\nu_a < \nu_m$ and $\nu_m < \nu_a$, respectively),   
 $\nu_{m, {\rm RS}} \propto \nu_{c,{\rm RS}} \propto t^{(15g+24)/(14g+7)}$, and $F_{\nu,{\rm max, RS}} \propto t^{-(11g+12)/(14g+7)}$.  In the case of a thick shell, $\nu_{m, {\rm RS}} \propto \nu_{c,{\rm RS}} \propto t^{-15/8}$ and $F_{\nu,{\rm max, RS}} \propto t^{-9/8}$, which in practice is almost identical to the Newtonian case for $g=1$.

Within our modeling, we follow the same procedure used with the forward shock by scaling in time backward after matching $\nu_{m, {\rm RS}}$ and $F_{\nu,{\rm max, RS}}$ against the well-determined $t=0.7$ day and $t \approx 2$ day SEDs in an attempt to reproduce the early evolution, with $g$ as a free parameter ($\nu_c$ is unimportant at $t>0.5$ day since the high frequencies are forward-shock dominated; its value is instead matched at early times).  The value of $p$ is assumed to be the same as in the forward shock (a reasonable assumption, since the shocks are initially in contact with each other).

There are two possible solutions, depending on the assumption of which break (peak) produces the fast switch from a hard spectrum to a soft one in the radio band at 0.7--2~d.  In the first case, this switch is caused by the minimum electron frequency $\nu_m$ passing through the band; this implies  $\nu_{a,{\rm RS}} < \nu_{m,{\rm RS}}$ and $g \approx 3.4$, and it is a reasonable fit to the data although it cannot explain the hard spectral index in the first C-band observation.  Alternatively, this could be caused by the synchrotron self-absorption break $\nu_a$, which implies  $\nu_{a,{\rm RS}} < \nu_{m,{\rm RS}}$ and $g \approx 3.0$.  (In either case $g=1$ is ruled out, implying the thin-shell case applies.)  Both cases predict generally identical behavior for $\nu > \nu_{\rm break}$, which means that only the earliest C-band point is capable of distinguishing them (and this frequency could be affected by radio scintillation, so the hard spectral index is not definitive).  However, for physical reasons (to be discussed in \ref{sec:interprevshock}) we prefer the second interpretation.

\subsection{Model Evaluation and Light Curves}

\begin{figure*}
\centerline{
\includegraphics[scale=0.85,angle=0]{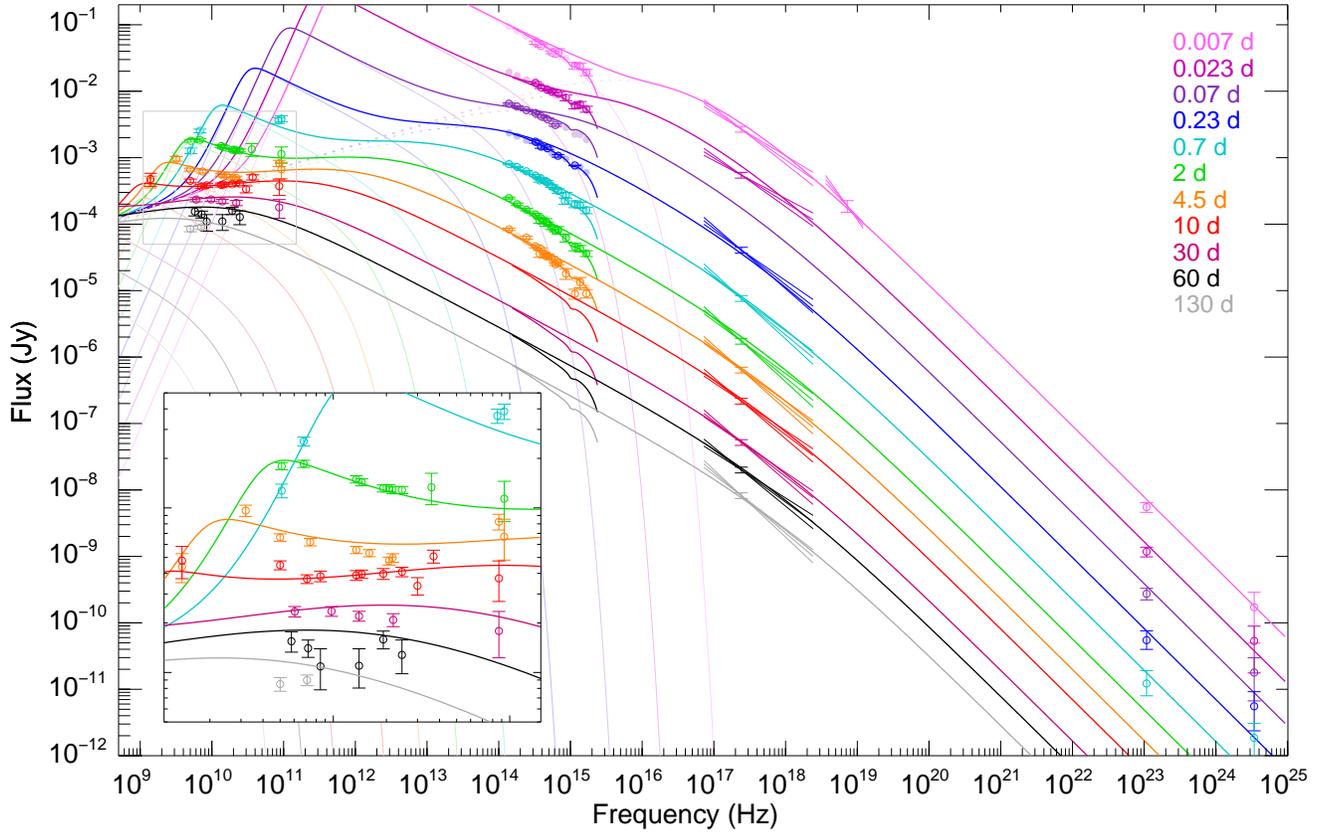}} 
\caption{Observations of the afterglow of GRB\,130427A spanning from the low-frequency radio to the 100 GeV LAT bands, interpolated to a series of coeval epochs spanning from 0.007~day (10~min) to 130 days after the burst.  Overplotted over each epoch is our simple forward+reverse shock model from standard synchrotron afterglow theory, which provides an excellent description of the entire dataset, a span of 18 orders of magnitude in frequency and 4 orders of magnitude in time.  The solid line shows the combined model, with the pale solid line showing the reverse-shock and the pale dotted line showing the forward-shock contribution.  The ``spur'' at $\approx10^{15}$~Hz shows the effects of host-galaxy extinction on the NIR/optical/UV bands.  Open points with error bars are measurements (adjusted to be coeval at each epoch time); pale filled points are model optical fluxes from the empirical fit in \S \ref{sec:optlc}.  The inset at lower left shows a magnified version of the radio part of the SED (gray box) at $t>0.7$ day.}
\label{fig:multised}
\end{figure*}

\begin{figure*}
\centerline{
\includegraphics[scale=0.85,angle=0]{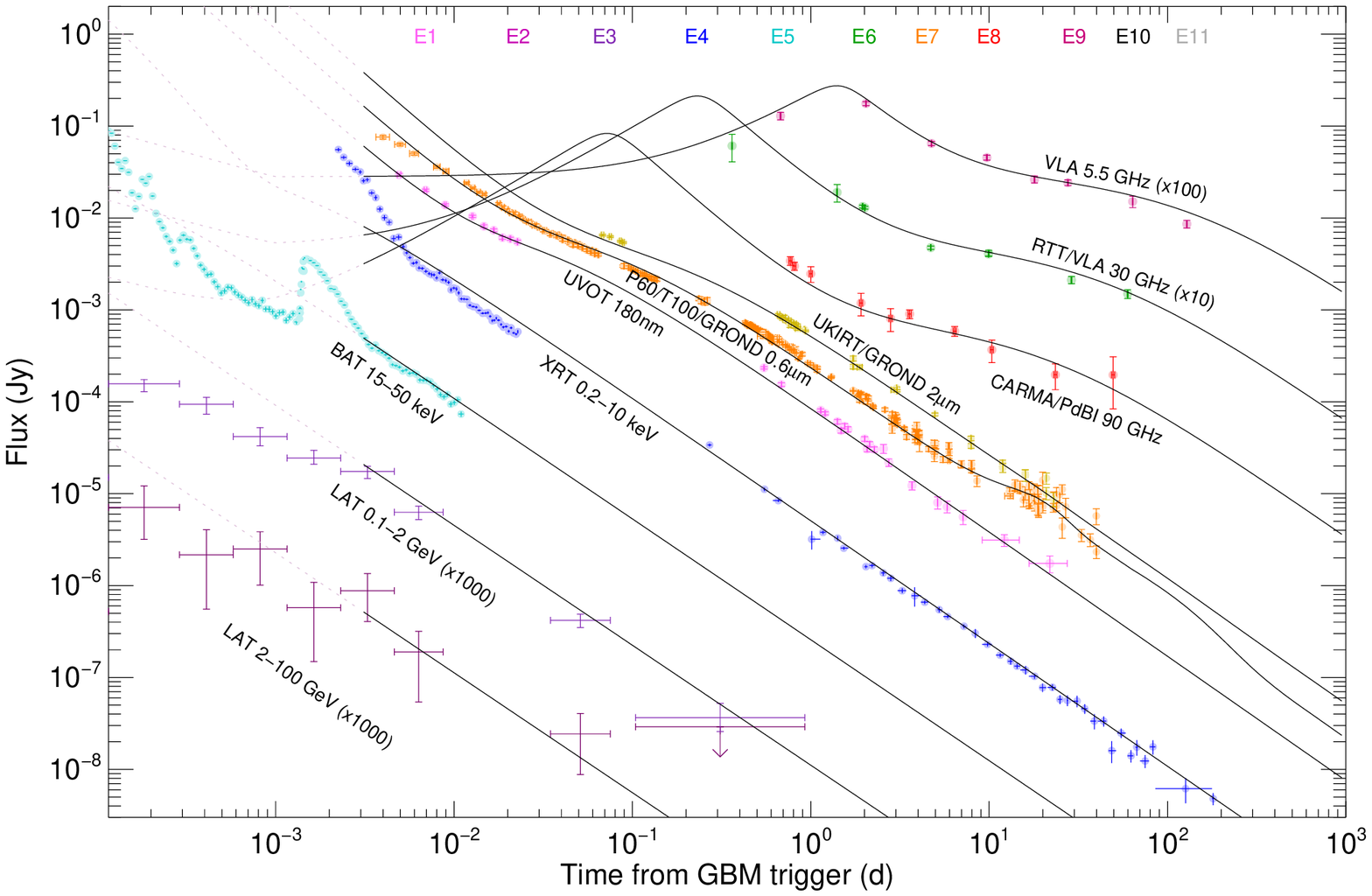}} 
\caption{Observed and analytic light curves of the afterglow of GRB\,130427A at specific frequencies: radio, millimeter, NIR, optical, UV, soft X-ray (XRT), hard X-ray (BAT), and extreme gamma-ray (LAT).  All of the major features at all frequencies are reproduced by our model (black lines), except at the earliest times.  The dotted lines show a naive extension of the model back in time, which generally overpredicts the fluxes at all frequencies (except during the final prompt-emission flare), perhaps due to the end of deceleration of the ejecta at these earliest epochs.  The numbers at the top indicate the times of the SED epochs shown in Figure \ref{fig:multised}.}
\label{fig:multilc}
\end{figure*}

This model (which has only a few significant free parameters in the form of the initial placement of the breaks, plus $p$, $g$, and the modified cooling-break index) does a remarkably good job fitting the data.  As can be seen in Figure \ref{fig:multised} the model curves are within about a factor of 2, and often much better, of each observation at all frequencies and at all epochs.   This is true for observations at intermediate epochs not shown in the discrete SEDs as well:  in Figure \ref{fig:multilc} we plot synthetic light curves from our model versus data at a series of frequencies spanning from the radio to high-energy gamma-rays.  All of the qualitative features of the GRB afterglow noted previously are represented by the model, including the relatively faint radio afterglow, the clear spectral index and steep-shallow-steep optical evolution, and the lack of X-ray spectral or temporal evolution.

The model is not perfect; in particular, it tends to underpredict the NIR flux slightly and implies a somewhat shorter-lived shallow-decay component in the optical bands relative to what is observed.  Nevertheless, given the only very limited flexibility in both the shape of the SED at any epoch and its evolution with time, the accuracy to which the observations are matched is remarkable.

\subsection{Physical Interpretation}
\label{sec:interpretation}

So far we have only taken the temporal \emph{scalings} of the break frequencies and peak fluxes from theory, with their initial values treated as free parameters.  However, these values are prescribed by theory as well, depending on the properties of the burst and its environment and microphysical parameters describing the partition of energy within the shock wave.  The parameters underlying the forward-shock evolution are $\epsilon_B$ (the fractional energy in magnetic fields), $\epsilon_e$ (the fractional energy in accelerated electrons), $E_K$ (the total energy of the shock), and the circumburst density normalization (in the case of a wind medium, $A_*$).  Additional fundamental parameters that are relevant to the reverse shock are $t_{\rm X}$ (the shock crossing time, which is $t_{\rm GRB}$ in the thick-shell model and $t_{\rm dec}$ in the thin-shell model), $\gamma_0$ (the initial Lorentz factor of the outflow, which is also the inverse baryon fraction of the initial fireball), and the magnetization $R_B$ (which permits $\epsilon_B$ in the ejecta to differ from that in the forward shock).  The electron index $p$, for which we have already solved, constitutes a final parameter.

\subsubsection{Forward Shock}
\label{sec:fwdshock}

To determine the forward-shock parameters we took the values of the spectral breaks and peak at the earliest epoch at which we have data ($t=0.007$ days) and inverted the standard equations governing the locations of the breaks to solve for $\epsilon_B$, $\epsilon_e$, $E_K$, and $A_*$.  (We choose this early epoch to try to minimize the impact of the subsequent noncanonical evolution of $\nu_c$.)   This solution is actually underdetermined because we could not measure $\nu_a$ directly, as it is below the observed bands at all times.  However, the physical requirement that $\epsilon_B + \epsilon_e < 1.0$ narrows down the allowed parameter space to a relatively narrow range.  We infer (assuming that neither efficiency exceeds its equipartition value; i.e., $\epsilon_B < 1/3$, $\epsilon_e < 1/3$)

$$0.03 < \epsilon_B < 0.33,$$
$$0.33  > \epsilon_e > 0.14,$$
$$1.9\times10^{53} < E_K < 4.2\times10^{53},$$
$$0.005 > A* > 0.001,$$

\noindent
where $E_K$ is in erg.  These ranges refer only to the range of ``best-fit'' parameters allowed given our constraints on $\epsilon_e$ (which determines the left bound) and $\epsilon_B$ (which determines the right bound).  Our inferred values are all consistent with the numbers presented by \cite{Laskar+2013} with the exception of $E_K$, which we find to be larger than their estimate.

The observed properties of the afterglow, in particular the radio faintness, can largely be explained as a product of the parameters observed for this GRB.  The large values of $E_K$ and $\epsilon_e$ and the low wind density $A*$ produce a shock with a great deal of energy distributed among a relatively small number of electrons, which therefore move very rapidly and radiate mostly at high frequencies (high $\nu_m$) at the expense of the lower-energy emission.  Specifically, $\nu_m$ is located in the optical band at early times, explaining why the afterglow appears blue at $t<0.5$ d but shifts to the red (and fades more rapidly) at later epochs.

\subsubsection{Reverse Shock}
\label{sec:interprevshock}

The observed properties of the reverse shock are determined by the same physical parameters as the forward shock with the addition of a direct dependence on the initial Lorentz factor $\gamma_0$ and the magnetization ratio $R_B = \epsilon_{B,{\rm RS}}/\epsilon_{B,{\rm FS}}$.  At the time of shell crossing ($\sim t_{\rm X}$), the values of $\nu_{m, {\rm RS}}$ and $F_{\nu,{\rm max}}$ are easily determined by simple scaling relations versus the equivalent values of the forward shock (for the thin-shell case, \citealt{Zou+2005}; and including the magnetization parameter as defined by \citealt{Gomboc+2008}):

$$F_{\nu,{\rm max,RS}}/F_{\nu,{\rm max,FS}} = 1.2 \gamma_0 R_B^{1/2},$$
$$\nu_{m,{\rm RS}}/\nu_{m,{\rm FS}} = 0.31 \gamma_0^{-2} R_B^{1/2},$$
$$\nu_{c,{\rm RS}}/\nu_{c,{\rm FS}} = R_B^{-3/2}.$$

While $\nu_{c,{\rm RS}}$ is generally hidden inside the forward shock at all times, $\nu_m$ and $F_{\nu, {\rm max,RS}}$ are tightly constrained observationally.  Based on their temporal scalings their values can then be extrapolated back to $t_X$, which is bounded by the burst duration ($t \approx 20$ s for the period of strongest emission) and the appearance of the afterglow in the BAT band ($t \approx 50$ s); this in turn can be used to solve for $\gamma_0$ and $R_B$.

As previously mentioned, there are two possible cases; $\nu_{a,{\rm RS}} < \nu_{m,{\rm RS}}$ and $\nu_{m,{\rm RS}} < \nu_{a,{\rm RS}}$.  The first possibility is strongly disfavored by this exercise as the Lorentz factor derived is extremely low ($\gamma_0 \approx 14$).  This would be in conflict with the independent constraint on $\gamma_0$ set by the deceleration time of the afterglow (which in the thin-shell case must be less than the afterglow peak time): e.g., from \cite{Zou+2005},

$$t_{\rm dec} = 2.9\times10^3 {\rm s} (1+z) E_{53} \gamma_{1.5}^{-4} A_{*,-1}^{-1}.$$
\noindent
For $t_{\rm dec} < 20$--50~s this would imply $\gamma_0 \gtrsim 120$--250, which is not consistent with the value set by the reverse-to-forward-shock ratio.  

We therefore adopt the $\nu_{m,{\rm RS}} < \nu_{a,{\rm RS}}$ model.  In this case, we find more reasonable values of $\gamma$ and $R_B$; e.g. for an assumed $t_X = 50$~s,

$$230 < \gamma_0 < 430,$$
$$2.3  < R_B  <  2.7,$$
\noindent
which is self-consistent with the deceleration constraint on $\gamma_0$.

A further consistency check can be performed by examining the location of the self-absorption break $\nu_{a,{\rm RS}}$, the peak value (at $t_X$) of which is predicted as a function of ($\epsilon_B$,$\epsilon_e$,$E_K$,$\Gamma$,$A_*$) by (for example) Equation 38 of \cite{Zou+2005}, and the time-evolution after that point is given by the power-law scalings discussed previously.  Its value at $t=0.7$d is well-constrained by observations ($\nu_{a, {\rm RS}}$ = 18 GHz); this value is fully consistent with its expected value from our theoretical model under the range of values derived above ($33$~GHz $ > \nu_{a,{\rm RS}} > 18$~GHz).

\subsection{Limits on the Collimation-Corrected Energetics}
\label{sec:energetics}

As seen in many other \Swift\ GRBs \citep{Racusin+2009}, the afterglow of GRB\,130427A shows no break in the XRT light curve in observations taken to date indicative of a jet.  Given that GRBs are believed to be narrowly collimated, this might initially seem surprising.  However, the jet break time is not just a function of beaming angle but also of other factors, including the energy of the shock and (in particular) the circumburst density.   The jet opening angle is related to these observable parameters by the relation \citep{LiChevalier2003}

$$ \theta_{\rm jet} = 5.4^{\circ} \left(\frac{t_{\rm jet}}{\rm d}\right)^{1/4} A_*^{1/4} \left(\frac{1+z}{2}\right)^{-1/4} \left(\frac{E_{\rm K}}{10^{53} {\rm ~erg}}\right)^{-1/4}. $$
\noindent
For the range of forward-shock parameters derived in \S \ref{sec:fwdshock}, we derive a constraint on the jet opening angle of $\theta_{\rm jet} > 5^{\circ}$ and a corresponding beaming-corrected $E_K > 5\times10^{50}$ erg and $E_\gamma > 2.2\times10^{51}$ erg.  These limits are still consistent with previous GRBs and the canonical energy scale of $\sim 10^{51}$~erg \citep{Frail+2001}, and the lack of a jet break through to the present time is no surprise --- in this case it is primarily a consequence of the very low density surrounding the burst.  In fact, a continued lack of a jet break out to extremely late times remains entirely consistent with the model; the burst is consistent with $E_K + E_{\gamma,{\rm iso}} < 10^{52}$ erg even for jet break times as late as 4~yr.

\section{Conclusions}
\label{sec:conclusions}

GRB\,130427A is a luminous and nearby, but otherwise remarkably ordinary, GRB. Its energy scale, while large, is well within the distribution of previously studied GRBs, before and (most likely) after correction for beaming.  As one of the closest, brightest, and best-observed GRBs to date and with no evidence of late-time flares, wide-angle jets, or energy re-injection events complicating its light curve, it offers an excellent study of afterglow evolution and a rigorous test of standard blastwave models.  

We attempted to fit the GRB afterglow properties as a combination of forward and reverse shocks evolving following the predictions of standard afterglow theory and, with the exception of nonstandard evolution of the cooling break, found extraordinarily good agreement at all times ($t\gtrsim400$~s) and frequencies ($10^9$~Hz $< \nu < 10^{25}$~Hz), in agreement with the results of \cite{Laskar+2013} for a much smaller dataset.   

All of the notable qualitative properties can be understood as a direct product of the particular fundamental burst parameters of this GRB, as follows.

The \emph{early fast optical and millimeter decay} is the product of a reverse shock initially peaking in the infrared and sweeping through to the millimeter and radio over the next several days.

The \emph{optical color changes} and shallow early-time $\beta_{\rm OX}$ can be understood as a result of the gradual transition from reverse to forward shock simultaneous with the movement of the forward-shock peak frequency through the optical band. 

The late-time \emph{radio faintness} is due to a combination of several factors, but can be qualitatively understood as the consequence of an energetic burst exploding into a very low-density, wind-swept environment, which concentrates the available shock energy into a small number of electrons and results in enhanced emission at higher frequencies at the expense of low frequencies.

The \emph{unbroken X-ray light curve} is a reflection of the canonical (with respect to basic theory) behavior of this burst: after 150~s there are no flares, refreshed shocks, central-engine wind, or other effects which would serve to boost the energy in the forward shock.   The lack of a jet break even out to late times most likely reflects the low density of its circumburst environment.

There is no reason to think that any of these properties are particularly unusual among the GRB population.  It has been noted in the past \citep{Schulze+2011} that relatively few GRBs show evidence of a wind-swept environment expected for a massive-star progenitor.  Our result for GRB\,130427A may show that this is, in part, a selection effect: robust constraints on the density profile require observations at radio wavelengths to track the evolution of $F_{\nu,{\rm peak}}$, but a wind-swept medium naturally suppresses this peak at late times and prevents detection of the afterglow.   The nonstandard evolution of the cooling break may also provide insight into this situation, as it would in particular negate the use of standard closure relations based solely on optical/X-ray data if $\nu > \nu_c$.

Among the parameters derived, the most remarkable is the very low wind density.  This requires a very low mass-loss rate; for a standard wind velocity of $10^3$~km~s$^{-1}$, our derived $A_*$ would indicate a mass-loss rate of only a few $\times 10^{-8}~{\rm M}_\odot$~yr$^{-1}$.  Mass-loss rates of this magnitude are a natural prediction for radiatively driven winds from massive, low-metallicity stars; for example, the modeling of \cite{Vink+2001} produces mass-loss rates below $10^{-7.5}~{\rm M}_\odot$~yr$^{-1}$ only for $Z < 0.05\,Z_\odot$.  Low mass-loss rates may also explain why density profiles typical of the interstellar medium are often preferred over wind-like ones; in a sufficiently dense environment this weak wind would clear out only a relatively small wind bubble \citep{vanMarle+2006}.   Low densities are not unprecedented, especially among very luminous GRBs: \cite{Cenko+2011} found similar, low values for a sample of four LAT-detected events from 2009.  With GRB\,130427A included, these results show clearly that low density is not rare and is no obstacle to the production of very high-energy gamma-rays, in contrast to some recent claims in the literature \citepeg{Beloborodov+2013}.  The apparent rarity of low-density, wind-driven environments among other GRB samples may be a selection effect; had more sensitive radio follow-up observations been more widely available in the past, similar signatures might have been observed more commonly, including among less luminous and more distant bursts.  The greatly improved sensitivity now available with the upgraded VLA and ALMA will soon test this prediction.

Our results also illustrate the value of early multi-frequency radio observations, especially at $t<1$ day.  Had the GRB been observed somewhat earlier ($t \approx 0.1$ d), even more dramatic evolution would have been observed; we predict that the millimeter light curve should have shown a rapid rise to a bright flare with a peak of 50 mJy between 15~min and 2~hr post-GRB.  The observation of such a signature would have presented even stronger verification of the reverse-shock interpretation for this GRB.  While such observations were not possible in this case, such a flare would be easily detectable even at significantly higher redshifts: owing to the steep slope below the self-absorption break, the $K$-correction during the flare rise is relatively favorable;  it would be detectable to CARMA and the VLA to at least $z \approx 1.2$ and to ALMA at almost any redshift.   A similar signature was previously seen in GRB 990123 \citep{Kulkarni+1999} and interpreted in a similar way; our results provide good reason to believe that this interpretation was correct and that this similar signature is probably ubiquitous among moderately luminous and nearby GRBs showing fast-decaying optical light curves at early times.

At the opposite end of the spectrum, the extrapolation of the forward-shock synchrotron SED to high frequency naturally explains the late-time GeV emission seen by the \textit{Fermi}-LAT for this and other bursts.  The long-lived nature of this emission has been a mystery since it was first hinted at by EGRET observations in the early 1990s \citep{Hurley+1994}.   While not completely precluding other possibilities, our observations provide strong support for the simplest possible explanation in this case, which is that it is primarily synchrotron emission from the forward shock \citepeg{Zou+2009,Gao+2009,Kumar+2009,Kumar+2010,Corsi+2010,dePasquale+2010,Ghirlanda+2010}.  Theoretically, synchrotron emission cannot easily produce photons at the very highest energies ($\gtrsim10$--100~GeV), and the detection of such photons probably requires an inverse-Comptonized contribution operating at the highest energies (observationally, there may be hints of an upturn in the SED in this range; \citealt{Fan+2013,Liu+2013}) --- but even in this case, the match in temporal and spectral properties with the lower-energy emission strongly ties it to the forward shock.

The success of our model in explaining the overall properties of this burst provides a strong vindication of the basic assumptions underlying standard GRB afterglow theory.  The range of behavior possible from standard afterglow theory is relatively limited and the opportunities for inconsistency were numerous, yet no irresolvable conflicts were encountered, and the parameters derived are in line with those observed from past GRBs and within reasonable expectation from theory.  While the profusion of data in the \Swift era produced innumerable examples of \emph{noncanonical} evolution of GRB afterglows, we show here that one of the most expansive datasets in time and frequency ever collected can still fit with good agreement to the standard theory with only very minor modifications.  This success greatly increases our confidence that the more complicated temporal and spectral evolution commonly seen in other GRBs with (flares, plateaus, rebrightenings) can indeed be understood by relatively simple extensions to the theory, such as energy input from a long-lived central engine wind, refreshed shocks, and wide-jet components.

\vskip 1cm

\acknowledgments

D.A.P. acknowledges a prescient conversation on 2013 April 18 with
A.~Collazzi in Nashville on the origin of the lack (at the time) of
{\it Fermi}-LAT GRBs at $z < 0.5$, and comments from Y.~Fan and 
A.~van~der~Horst.   We are grateful for excellent
staff assistance at the various observatories where we obtained data.
We thank B.~Sesar for the execution of our P200 target-of-opportunity
observations and assistance with the data reduction. J. Carpenter and
CARMA provided additional DDT time to continue observations of GRB
130427A after $t=2$ days, and K.~Alatalo assisted with flagging and
reducing these observations.  We thank S.~Kulkarni for the Keck/LRIS
data.  We would like to thank all members of the \emph{Swift} and 
\emph{Fermi} teams, whose successful missions made this work possible.
  The National Radio Astronomy Observatory is a facility of the
NSF operated under cooperative agreement by Associated Universities,
Inc.  Support for CARMA construction was derived from the states of
California, Illinois, and Maryland, the James S. McDonnell Foundation,
the Gordon and Betty Moore Foundation, the Kenneth T. and Eileen
L. Norris Foundation, the University of Chicago, the Associates of the
California Institute of Technology, and the NSF. Ongoing CARMA
development and operations are supported by the NSF under a
cooperative agreement, and by the CARMA partner universities. 
 The article contains data taken with the IRAM Plateau
deBure interferometer. IRAM is supported by INSU/CNRS (France), MPG
(Germany) and IGN (Spain). We acknowledge the support of the Spanish
Ministerio de Ciencia y Tecnolog\'{\i}a through Projects
AYA2009-14000-C03-01/ESP and AYA2012-39727-C03-01.  
KAIT and its ongoing operation were made possible by donations from Sun
Microsystems, Inc., the Hewlett-Packard Company, AutoScope
Corporation, Lick Observatory, the NSF, the University of California,
the Sylvia and Jim Katzman Foundation, and the TABASGO
Foundation. Some of the data presented herein were obtained at the
W. M. Keck Observatory, which is operated as a scientific partnership
among the California Institute of Technology, the University of
California, and NASA; the Observatory was made possible by the
generous financial support of the W. M. Keck Foundation.  This work
made use of data supplied by the UK {\it Swift} Science Data Centre at
the University of Leicester, and the NASA Extragalactic Database.
Part of the funding for GROND (both hardware as well as personnel) 
was generously granted from the Leibniz-Prize to Prof. G. Hasinger 
(DFG grant HA 1850/28-1).  D.A.K. acknowledges F. Ludwig and U. Laux 
for helping to obtain the TLS images. 

Support for this research was provided by NASA through Hubble
Fellowship grant HST-HF-51296.01-A awarded by the Space Telescope
Science Institute, which is operated for NASA by the Association of
Universities for Research in Astronomy, Inc., under contract NAS
5-26555.
A.V.F.'s group at UC Berkeley has
received generous financial assistance from Gary and Cynthia Bengier,
the Christopher R. Redlich Fund, the Richard and Rhoda Goldman Fund,
the TABASGO Foundation, NSF grant AST-1211916, and NASA/{\it Swift}
grants NNX10AI21G and NNX12AD73G.  E. S. acknowledges assistance from
the Scientific and Technological Research Council of Turkey 
(T\"UB\.ITAK) through project 112T224.
 X.H.Z. acknowledges partial support by the NSFC
(No. 11203067), Yunnan Natural Science Foundation (2011FB115), and the
West Light Foundation of the CAS. 
J. Mao is supported by Grants-in-Aid for Foreign JSPS Fellow (No. 24.02022).
J. M. Bai is supported by the NSFC (No. 11133006).
A.S.P. and A.A.V. acknowledge assistance from RFFI grants 12-02-01336 and 13-01-92204.
S.S. acknowledges support by the 
Th\"uringer Ministerium f\"ur Bildung, Wissenschaft und Kultur under 
FKZ 12010-514. S.K. acknowledges support by DFG grant Kl 766/16-1.

\bigskip 

{\it Facilities:} \facility{Swift}, \facility{Fermi}, \facility{PO:1.5m}, \facility{UKIRT}, \facility{Max Planck:2.2m}, \facility{PO:Hale}, \facility{Gemini:Gillett}, \facility{CARMA}, \facility{VLA}, \facility{IRAM:PdBI}, \facility{Keck:I}, \facility{Lick:Shane}, \facility{Lick:KAIT}, \facility{Lick:Nickel}


\begin{deluxetable}{llcccll}
\tabletypesize{\small}
\tablecaption{Photometry of GRB\,130427A\tablenotemark{*}}
\tablecolumns{7}
\tablehead{
\colhead{Telescope} &
\colhead{$t$\tablenotemark{a}} &
\colhead{Filter} &
\colhead{Exp.~time} &
\colhead{Mag.\tablenotemark{b}} & 
\colhead{Flux\tablenotemark{c}}  \\
\colhead{} &
\colhead{(day)} & \colhead{} &
\colhead{(s)} & \colhead{} &
\colhead{($\mu$Jy)} }
\startdata
UVOT       &   0.00492 & UVW2                  &   19.5 & $11.044 \pm 0.040$ & $   29499\pm    1067$ \\
UVOT       &   0.00520 & UVV                   &   19.5 & $12.169 \pm 0.030$ & $   53489\pm    1458$ \\
UVOT       &   0.00549 & UVM2                  &   19.5 & $11.158 \pm 0.030$ & $   31324\pm     854$ \\
UVOT       &   0.00577 & UVW1                  &   19.4 & $11.228 \pm 0.040$ & $   32560\pm    1178$ \\
UVOT       &   0.00836 & UVB                   &   19.5 & $12.899 \pm 0.160$ & $   31093\pm    4261$ \\
UVOT       &   0.01006 & UVU                   &   19.4 & $12.177 \pm 0.040$ & $   20863\pm     755$ \\
P60        &   0.00400 & r                     &     60 & $11.751 \pm 0.053$ & $   75792\pm    3811$ \\
P60        &   0.00499 & i                     &     60 & $11.808 \pm 0.041$ & $   71012\pm    2758$ \\
P60        &   0.00598 & z                     &     60 & $11.928 \pm 0.040$ & $   64485\pm    2447$ \\
P60        &   0.09069 & g                     &     60 & $15.534 \pm 0.035$ & $    2367\pm      75$ \\
UKIRT      &   0.06799 & K                     &    180 & $12.530 \pm 0.020$ & $    6509\pm     119$ \\
UKIRT      &   0.07875 & J                     &    180 & $13.839 \pm 0.020$ & $    4692\pm      86$ \\
UKIRT      &   0.08575 & H                     &    180 & $13.294 \pm 0.020$ & $    4952\pm      91$ \\
GMG        &   0.23591 & r                     &    180 & $16.135 \pm 0.057$ & $    1325\pm      68$ \\
GMG        &   0.24050 & B                     &    180 & $16.610 \pm 0.040$ & $    1005\pm      37$ \\
GMG        &   0.24317 & z                     &    180 & $15.941 \pm 0.060$ & $    1586\pm      86$ \\
GMG        &   0.24557 & i                     &    180 & $16.086 \pm 0.049$ & $    1368\pm      61$ \\
T100       &   0.42937 & R                     &     60 & $16.555 \pm 0.050$ & $     762\pm      35$ \\
GROND      &   0.66852 & g                     & 2112.5 & $17.544 \pm 0.042$ & $     366\pm      14$ \\
GROND      &   0.65651 & r                     & 1056.2 & $17.283 \pm 0.049$ & $     452\pm      21$ \\
GROND      &   0.65651 & i                     & 1056.2 & $17.052 \pm 0.040$ & $     555\pm      21$ \\
GROND      &   0.65651 & z                     & 1056.2 & $16.989 \pm 0.034$ & $     595\pm      19$ \\
GROND      &   0.65898 & J                     &   1200 & $15.887 \pm 0.030$ & $     697\pm      19$ \\
GROND      &   0.65898 & H                     &   1200 & $15.273 \pm 0.031$ & $     784\pm      23$ \\
GROND      &   0.68055 & K                     &    960 & $14.736 \pm 0.053$ & $     837\pm      41$ \\
KAIT       &   0.84703 & V                     &     -- & $17.810 \pm 0.040$ & $     283\pm      11$ \\
KAIT       &   0.84877 & R                     &     -- & $17.490 \pm 0.040$ & $     315\pm      12$ \\
KAIT       &   0.85333 & B                     &     -- & $18.270 \pm 0.120$ & $     214\pm      23$ \\
KAIT       &   0.87987 & I                     &     -- & $17.060 \pm 0.100$ & $     363\pm      33$ \\
Nickel     &   0.95267 & B                     &    300 & $18.241 \pm 0.057$ & $     220\pm      11$ \\
Nickel     &   0.95807 & V                     &    200 & $17.910 \pm 0.024$ & $     258\pm       6$ \\
Nickel     &   0.96136 & R                     &    150 & $17.632 \pm 0.027$ & $     275\pm       7$ \\
Nickel     &   0.96371 & I                     &    120 & $17.257 \pm 0.045$ & $     301\pm      13$ \\
...        &   ...     & ...                   &   ...  & ...                & ...               \\
\enddata
\tablenotetext{*}{Only the first exposure for each telescope and filter (excluding $t>1$ day) is shown.   A full table containing all 707 data points will be published online and is also available at \url{http://www.astro.caltech.edu/~dperley/grb/130427a/data/}.}
\tablenotetext{a}{Exposure mid-time, measured from the \emph{Fermi}-GBM trigger (UT 07:47:06.42).}
\tablenotetext{b}{Observed value, not corrected for Galactic extinction and including host-galaxy flux.}
\tablenotetext{c}{Corrected for Galactic extinction ($E_{B-V} = 0.02$ mag) and host-galaxy contribution.  Except for the P60 observations, uncertainties do not include the uncertainty resulting from subtraction of the host flux; this is negligible over most of the afterglow evolution but contributes a large, systematic uncertainty during the SN phase.}
\label{tab:optphot}
\end{deluxetable}

\begin{deluxetable}{llcccl}
\tabletypesize{\small}
\tablecaption{Radio Observations of GRB\,130427A}
\tablecolumns{6}
\tablehead{
\colhead{Telescope} &
\colhead{$t$\tablenotemark{a}} &
\colhead{Band} &
\colhead{Frequency\tablenotemark{b}} &
\colhead{Flux} \\
\colhead{} &
\colhead{(day)} &
\colhead{} &
\colhead{(GHz)} &
\colhead{($\mu$Jy)} }
\startdata
RT22    &   0.36173 & Ka  &  36.00 & $ 6100\pm 2023$ \\
VLA     &   0.67749 & C   &   5.10 & $ 1290\pm  123$ \\
VLA     &   0.67749 & C   &   6.80 & $ 2570\pm  160$ \\
CARMA   &   0.76900 & 3mm &  93.00 & $ 3416\pm  365$ \\
CARMA\tablenotemark{c}   &   0.81000 & 3mm &  85.00 & $ 3000\pm  300$ \\
CARMA   &   1.00000 & 3mm &  93.00 & $ 2470\pm  481$ \\
RT22    &   1.40524 & Ka  &  36.00 & $ 1900\pm  411$ \\
CARMA   &   1.91400 & 3mm &  93.00 & $ 1189\pm  327$ \\
VLA     &   1.95438 & K   &  20.70 & $ 1349\pm   69$ \\
VLA     &   1.95438 & K   &  21.70 & $ 1314\pm   67$ \\
VLA\tablenotemark{c}     &   2.00877 & K   &  19.20 & $ 1310\pm   65$ \\
VLA\tablenotemark{c}     &   2.00877 & K   &  24.50 & $ 1280\pm   64$ \\
VLA\tablenotemark{c}     &   2.01811 & Ku  &  13.50 & $ 1480\pm   74$ \\
VLA\tablenotemark{c}     &   2.01811 & Ku  &  14.50 & $ 1420\pm   71$ \\
VLA\tablenotemark{c}     &   2.03854 & C   &   5.10 & $ 1760\pm   88$ \\
VLA\tablenotemark{c}     &   2.03854 & C   &   6.80 & $ 1820\pm   91$ \\
CARMA   &   2.80300 & 3mm &  93.00 & $  807\pm  225$ \\
GMRT\tablenotemark{c}    &   3.25000 & L   &   1.39 & $  500\pm  100$ \\
PdBI    &   3.58256 & 3mm &  86.74 & $  903\pm   99$ \\
VLA     &   4.71239 & K   &  20.70 & $  467\pm   26$ \\
VLA     &   4.71239 & K   &  21.70 & $  488\pm   29$ \\
VLA     &   4.73196 & U   &  13.50 & $  545\pm   28$ \\
VLA     &   4.73196 & U   &  16.00 & $  521\pm   27$ \\
VLA     &   4.74985 & C   &   5.00 & $  648\pm   34$ \\
VLA     &   4.74985 & C   &   7.40 & $  607\pm   32$ \\
VLA     &   4.76347 & S   &   3.19 & $  942\pm   74$ \\
PdBI    &   6.41243 & 3mm &  86.74 & $  587\pm   71$ \\
VLA     &   9.71473 & C   &   5.00 & $  454\pm   28$ \\
VLA     &   9.71473 & C   &   7.10 & $  374\pm   23$ \\
VLA     &   9.92480 & K   &  19.20 & $  399\pm   30$ \\
VLA     &   9.92480 & K   &  24.50 & $  410\pm   27$ \\
VLA     &   9.93576 & Ka  &  30.00 & $  337\pm   41$ \\
VLA     &   9.93576 & Ka  &  37.00 & $  509\pm   46$ \\
VLA     &   9.94673 & U   &  13.50 & $  390\pm   27$ \\
VLA     &   9.94673 & U   &  14.50 & $  397\pm   23$ \\
VLA     &   9.95608 & X   &   8.46 & $  385\pm   28$ \\
PdBI    &  10.35652 & 3mm &  86.74 & $  368\pm  102$ \\
GMRT\tablenotemark{c}    &  11.60000 & L   &   1.39 & $  450\pm  100$ \\
VLA     &  17.91450 & U   &  14.00 & $  275\pm   31$ \\
VLA     &  17.92385 & X   &   8.50 & $  332\pm   72$ \\
VLA     &  17.95399 & C   &   5.50 & $  263\pm   22$ \\
PdBI    &  23.51729 & 3mm &  86.74 & $  197\pm   62$ \\
VLA     &  27.65017 & C   &   6.05 & $  242\pm   18$ \\
VLA     &  27.66800 & X   &   9.77 & $  243\pm   16$ \\
VLA     &  28.93510 & K   &  21.85 & $  212\pm   19$ \\
VLA     &  28.95414 & U   &  14.00 & $  223\pm   15$ \\
PdBI    &  49.56798 & 3mm &  86.74 & $  196\pm  112$ \\
VLA     &  59.78386 & K   &  19.20 & $  159\pm   20$ \\
VLA     &  59.78386 & K   &  24.50 & $  128\pm   30$ \\
VLA     &  59.79915 & U   &  14.00 & $  110\pm   30$ \\
VLA     &  59.80989 & X   &   8.46 & $  109\pm   30$ \\
VLA     &  63.77799 & C   &   7.20 & $  137\pm   16$ \\
VLA     &  63.77799 & C   &   5.79 & $  151\pm   21$ \\
VLA     &  63.78838 & S   &   3.10 & $  119\pm   45$ \\
VLA     & 128.33868 & C   &   5.00 & $   86\pm    8$ \\
VLA     & 128.33868 & C   &   7.10 & $   91\pm    7$ \\
\enddata 
\tablenotetext{a}{Integration mid-time, measured from the \emph{Fermi}-GBM trigger (UT 07:47:06.42).}
\tablenotetext{b}{Central frequency.}
\tablenotetext{c}{From Laskar et al.\ 2013.}
\label{tab:radiophot}
\end{deluxetable}

\begin{deluxetable}{lll}
\tabletypesize{\small}
\tablecaption{Host-Galaxy Magnitudes}
\tablecolumns{3}
\tablehead{
\colhead{Filter} &
\colhead{Mag\tablenotemark{a}} &
\colhead{AB mag} }
\startdata
UVW2 & 21.30 & 23.14 \\
UVM2 & 21.25 & 22.94 \\
UVW1 & 21.10 & 22.62 \\
UVU  & 21.60 & 22.62 \\
B    & 22.43 & 22.29 \\
g    & 21.98 & 21.98 \\
V    & 21.49 & 21.47 \\
r    & 21.26 & 21.26 \\
R    & 21.06 & 21.23 \\
i    & 21.19 & 21.19 \\
I    & 20.71 & 21.15 \\
z    & 21.03 & 21.01 \\
J    & 19.94 & 20.84 \\
H    & 19.35 & 20.73 \\
K    & 18.90 & 20.74 \\
\enddata
\tablenotetext{a}{Magnitude assumed when subtracting from the observed fluxes to isolate the afterglow/supernova.  Magnitudes are in the standard system for each instrument \citep{Poole+2008,Fukugita+1996,Cohen+2003} and are not corrected for Galactic extinction.}
\label{tab:host}
\end{deluxetable}

\begin{deluxetable}{lll}
\tabletypesize{\small}
\tablecaption{Coeval SEDs}
\tablecolumns{3}
\tablehead{
\colhead{Frequency} &
\colhead{Flux} &
\colhead{Uncertainty} \\
\colhead{(Hz)} &
\colhead{($\mu$Jy)}  &
\colhead{($\mu$Jy)} }
\startdata
\hline
\multicolumn{3}{l}{ t = 0.007 day } \\
3.251e+14 &        53802 &         2967 \\
3.890e+14 &        47988 &         2263 \\
5.483e+14 &        38389 &         1705 \\
6.826e+14 &        38241 &         5459 \\
1.153e+15 &        24285 &         2793 \\
1.363e+15 &        23840 &         2428 \\
1.666e+15 &        19216 &         2118 \\
2.418e+17 &         2995 &          299 \\
7.254e+18 &          188 &           38 \\
1.081e+23 &      5.52e-03 &      9.20e-04 \\
3.420e+24 &      1.72e-04 &      1.16e-04 \\
\hline
\multicolumn{3}{l}{ t = 0.023 day } \\
3.251e+14 &        13356 &          605 \\
3.890e+14 &        11499 &          496 \\
4.766e+14 &        10479 &          438 \\
5.483e+14 &         9955 &          509 \\
6.826e+14 &         9157 &          500 \\
8.652e+14 &         8006 &          815 \\
1.153e+15 &         6063 &          626 \\
1.363e+15 &         6337 &          671 \\
1.666e+15 &         5386 &          556 \\
2.418e+17 &          547 &           55 \\
1.081e+23 &      1.18e-03 &      1.97e-04 \\
3.420e+24 &      5.34e-05 &      3.58e-05 \\
\hline
\multicolumn{3}{l}{ t = 0.07 day } \\
...       &       ...     &       ...     \\
\enddata
\tablenotetext{*}{Only the first two epochs are shown.   A full table containing all 11 epochs will be published online and is also available at \url{http://www.astro.caltech.edu/~dperley/grb/130427a/data/}.}
\label{tab:seds}
\end{deluxetable}

\end{document}